\documentclass{elsart}

\usepackage{natbib}

\usepackage{graphicx}
\usepackage{subfigure}

\usepackage{amssymb}

\begin{document}

\begin{frontmatter}



\title{The pulsing CPSD method for subcritical assemblies with pulsed sources}


\author{Daniel Ballester\corauthref{cor1}} and
\ead{dabalber@mat.upv.es} \corauth[cor1]{Corresponding author: Fax +34 963 877 669}
\address{Department of Applied Mathematics,\\ Polytechnic University of Valencia,
46022 Valencia, Spain}
\author{Jos\'e L. Mu\~noz-Cobo}
\ead{jlcobos@iqn.upv.es}
\address{Department of Chemical and Nuclear Engineering,\\ Polytechnic University of Valencia,
46022 Valencia, Spain}

\begin{abstract}
Stochastic neutron transport theory is applied to the derivation of the
two-neutron-detectors cross power spectral density for subcritical assemblies when
external pulsed sources are used. A general relationship between the two-detector probability
generating functions of the kernel and the source is obtained considering the
contribution to detectors statistics of both the pulsed source and the intrinsic neutron
source. An expansion in $\alpha$-eigenvalues is derived for the final solution, which
permits to take into account the effect of higher harmonics in subcritical systems.
Further, expressions corresponding to the fundamental mode approximation are compared
with recent results from experiments performed under the MUSE-4 European research
project.
\end{abstract}




\end{frontmatter}


\section{Introduction}

In last years, researchers have shown an increasing interest on the conceptual
development of accelerator-driven systems (ADS) for nuclear waste transmutation and
energy production purposes. An important issue regarding its future industrial
applicability is the development of a periodically subcriticality level measurement and
monitoring technique, since both its operation safety and its performance as a part of
the nuclear fuel cycle shall be seriously affected by this variable.

Following the study of the neutron fluctuations in a multiplying medium, \citet{Courant},
several static and dynamic methods have been proposed and studied for years concerning
their applicability for the determination of some nuclear reactor physics
parameters (see \citet{Uhrig,Williams74,Lewins78,Carta99}). Within the group of dynamic
techniques, apparently the utilisation of the neutron-fission chain fluctuations for
nuclear assemblies subcriticality determination was firstly suggested by Bruno Rossi (the
Rossi-$\alpha $ method). In these methods neutron detector counting rates related to
individual fission-chain events must be discerned from the total counting rate, therefore
these methods are applicable to subcritical systems near delayed critical
conditions, \citet{Carta99}.

Further, dynamic methods based on a time-dependent external neutron source
were proposed by \citet{Perez65}. Recently, the use of these methods has increased during the
MUSE European experimental studies carried out at the MASURCA facility (Cadarache,
France) due to their applicability in order to investigate ADS kinetic parameters. In
these experiments D-D and D-T neutron sources running in pulsed mode have been used,
although the utilisation of an external spallation proton source has also been thought.
Anyway, measurements can be done in two different
ways (\citet{Valentine2000,Degweker2003,Ceder2003}): in the first one, the neutron detector
time gate is synchronized with the external neutron pulse injection, thus this method is
referred to as deterministic pulsing method, whereas, in the second case, the relative
delay between the neutron pulse injection and the beginning of the neutron counting time
is uniformly sampled between zero and the pulsed source period, which is known as
stochastic pulsing method. For the latter case, the neutron source can be assumed to have
the form%

\begin{equation}
S\left( t\right) = k \sum_{m=-\infty }^{\infty }\delta \left( t-\left( \xi +mT\right) \right) ,
\label{spsource}
\end{equation} where $k$ is the number of protons injected per proton pulse, $T$ is the
pulsed source period, and $\xi $ is uniformly sampled within the time interval $[0,T]$.
Obviously, for the deterministic pulsing method we will put $\xi =0$.

On the other hand, \citet{Pal58} developed a general theory for the study of the stochastic
neutron field. This model, complemented afterwards by \citet{Bell65},
completely describe the stochastic neutron field in a fissile assembly. Later, \citet{MPV87} derived expressions for the variance of the number of counts in a detector
and the CPSD for a Poissonian source without delayed neutrons from the general neutron
stochastic transport theory. These approaches permit to go beyond classical
point kinetic approximations, taking into consideration general problems with spatial,
spectral, and angular dependence. They have been extensively applied to nuclear
subcriticality safety and non-destructive nuclear fuel assay problems.

Classical reactor noise methods are not correct when pulsed or correlated sources are
used (\citet{Matthes88,Behringer99}). For these kind of sources, forward Kolmogorov's
approach is not valid because of the non-Markovian character of the process, while
backward Green's function description needs to go beyond Poissonian
behaviour (\citet{Degweker2003,BM2005}).

In our work we have derived a relationship between the source probability generating
function and the kernel probability generating function when non-Poissonian spallation
neutron sources are considered. We have also studied the effect of the intrinsic neutron
source due to spontaneous fission occurring in major actinides forming part of the
nuclear fuel of an ADS. This result can considered as a generalisation of the master
equation obtained by \citet{BM2005} for cross statistical descriptors (problems
with more than one detector).

In particular, we admit the intrinsic source spontaneous fission process to behave as a
Poissonian one, albeit neutron emission multiplicity corresponding to spontaneous fission
events has also been included.

In this paper we have neglected the contribution of delayed neutrons, therefore all
quantities appearing here can be interpreted as prompt variables for short time-scales,
in comparison with the delayed neutron precursors lifetimes.


\section{General expression for the relationship between the source pgf and the kernel pgf}\label{general}

The derivation of the general relationship between the source pgf and the kernel pgf
shall be based on two well known results (see, e.g., \citet{Lando}):
\begin{itemize}
\item given a random variable, $\mathcal{Z=X}_{1}+\mathcal{X}_{2}+\ldots +\mathcal{X}_{n}$,
$\mathcal{X}_{i}$, $i=1,2,\ldots ,n$, being mutually independent discrete random
variables, and $G_{\mathcal{X}_{i}}(s)=\sum_{j}s^{j}P_{\mathcal{X}_{i}=j} $ being the
probability generating function associated to $\mathcal{X}_{i}$, where
$P_{\mathcal{X}_{i}=j} $ is the probability for the occurrence of the event
$\mathcal{X}_{i}=j$, $\sum_{j}P_{\mathcal{X}_{i}=j}=1$, then the probability generating
function of $\mathcal{Z}$ can be expressed as

\begin{equation}
G_{\mathcal{Z}}\left( s\right) =\prod_{i}G_{\mathcal{X}_{i}}\left( s\right) ;  \label{apgzet}
\end{equation}

\item given a random variable $\mathcal{Z}=\mathcal{X}_{1}+\mathcal{X}_{2}+\ldots +
\mathcal{X}_{\mathcal{Y}}$, $\mathcal{X} \equiv \mathcal{X}_{i} $ $\forall i$ and
$\mathcal{Y}$ being mutually independent discrete random variables, then

\begin{equation}
G_{\mathcal{Z}}\left( s\right) =G_{\mathcal{Y}}\left( G_{\mathcal{X}}\left(s\right) \right) .  \label{apgz}
\end{equation}

\end{itemize}

In our system we must consider two independent sources leading to detector
counts, \citet{BM2005}: neutrons appearing from spontaneous fission of isotopes contained
within the nuclear fuel, and neutrons coming from the spallation source induced by
nuclear interactions of the proton beam with the target material.

In particular, for the derivation of the cross correlation between two neutron detectors,
we shall consider the two-dimensional discrete random variable $\mathcal{Z} =
\left(\mathcal{Z}_{1},\mathcal{Z}_{2}\right)$, where $\mathcal{Z}_{i}$, $i=1,2$, is the
number of neutron detections gathered by the $i$-th detector during the time interval
$\left( t_{{\rm f}_{i}}-\tau_{{\rm c}_{i}},t_{{\rm f}_{i}}\right) $. And according to our
previous discussion, $\mathcal{Z}_{i}=\mathcal{N}_{i}+\mathcal{M}_{i}$, where
$\mathcal{N}_{i}$, $\mathcal{M}_{i}$, are the number of detections registered by the
$i$-th detector and coming from the intrinsic spontaneous fission source and the external
pulsed source, respectively; equivalently, $\mathcal{N} =
\left(\mathcal{N}_{1},\mathcal{N}_{2}\right)$, $\mathcal{M} =
\left(\mathcal{M}_{1},\mathcal{M}_{2}\right)$.

In case of the intrinsic spontaneous fissions source, the number of detector counts
within the time interval $\left( t_{{\rm f}_{i}}-\tau_{{\rm c}_{i}},t_{{\rm
f}_{i}}\right) $, $\mathcal{N}_{i}$, can be expressed as

\begin{equation}
\mathcal{N}_{i}=\mathcal{X}_{i_{1}}+\mathcal{X}_{i_{2}}+\ldots
+ \mathcal{X}_{i_{\mathcal{Y}}} , \label{NXY}
\end{equation} where $\mathcal{Y}$ is the number of spontaneous fission events occurring
within the fuel material and $\mathcal{X}_{i_{j}} \equiv \mathcal{X}_{i} $ is the number of detector counts gathered
by the $i$-th detector corresponding to each spontaneous fission event, both being
mutually independent discrete random variables, \citet{BM2005}; its corresponding
two-dimensional discrete random variable will be denoted as $\mathcal{X} =
\left(\mathcal{X}_{1},\mathcal{X}_{2}\right)$. We can define the following probability
functions: $P_{n_{1} n_{2}} \left( d_{1}\left( t_{{\rm f}_{1}}\right),d_{2}\left(t_{{\rm
f}_{2}}\right) \right)$ is the joint probability to have $\mathcal{N}_{i}=n_{i}$,
$i=1,2$, detector counts within the time interval $\left( t_{{\rm f}_{i}}-\tau_{{\rm
c}_{i}},t_{{\rm f}_{i}}\right) $ upon the introduction of the neutron intrinsic source in
the remote past, $P_{y} \left( t_{\rm f} \right)$ is the probability to have
$\mathcal{Y}=y$ intrinsic source spontaneous fission events within the fuel material at
time $t_{\rm f}=\max\left\{ t_{{\rm f}_{1}},t_{{\rm f}_{2}}\right \}$ upon the
introduction of this neutron source in the remote past, $P_{x_{1} x_{2}} \left(
\mathbf{r},t|d_{1}\left( t_{{\rm f}_{1}}\right),d_{2}\left(t_{{\rm f}_{2}}\right)\right)$
is the joint probability to have $\mathcal{X}_{i}=x_{i}$, $i=1,2$, counts in the $i$-th
detector within the time interval $\left( t_{{\rm f}_{i}}-\tau_{{\rm c}_{i}},t_{{\rm
f}_{i}}\right) $ after an intrinsic source disintegration event at time $t$ and position
$\mathbf{r}$. In addition, $K_{n_{1}n_{2}}\left( \vartheta, t | d_{1}\left( t_{{\rm
f}_{1}}\right),d_{2}\left(t_{{\rm f}_{2}}\right) \right)$ is the joint probability of
having $\mathcal{N}_{i}=n_{i}$, $i=1,2$, detector counts per single neutron injected in
the phase-space point $\vartheta=\left( \mathbf{r},v,\mathbf{\Omega} \right)$ at instant
$t$. Associated with these probabilities we have the source probability generating
functions

\begin{equation}
G_{\rm S}^{\rm sf}\left( s_{1},s_{2}|d_{1}\left( t_{{\rm
f}_{1}}\right),d_{2}\left(t_{{\rm f}_{2}}\right) \right) = \sum_{n_{1},n_{2}=0}^{\infty}
s_{1}^{n_{1}} s_{2}^{n_{2}} P_{n_{1}n_{2}} \left(d_{1}\left( t_{{\rm
f}_{1}}\right),d_{2}\left(t_{{\rm f}_{2}}\right)\right), \label{GSsf}
\end{equation}

\begin{equation}
G_{\rm S,\mathcal{Y}}\left( s, t_{\rm f}\right)= \sum_{y=0}^{\infty} s^{y} P_{y} \left(
t_{\rm f} \right), \label{GSY}
\end{equation}

\begin{equation}
G_{\rm S,\mathcal{X}}\left( s_{1},s_{2},\mathbf{r},t|d_{1}\left( t_{{\rm
f}_{1}}\right),d_{2}\left(t_{{\rm f}_{2}} \right) \right) = \sum_{x_{1},x_{2}=0}^{\infty}
s_{1}^{x_{1}} s_{2}^{x_{2}} P_{x_{1}x_{2}} \left( \mathbf{r},t|d_{1}\left( t_{{\rm
f}_{1}}\right),d_{2}\left(t_{{\rm f}_{2}}\right)\right), \label{GSX}
\end{equation} and the kernel probability generating function

\begin{equation}
G_{\rm K}\left(s_{1},s_{2}, \vartheta ,t|d_{1}\left( t_{{\rm
f}_{1}}\right),d_{2}\left(t_{{\rm f}_{2}}\right) \right) = \sum_{n_{1},n_{2}=0}^{\infty}
s_{1}^{n_{1}}s_{2}^{n_{2}} K_{n_{1}n_{2}}\left( \vartheta, t |d_{1}\left( t_{{\rm
f}_{1}}\right),d_{2}\left(t_{{\rm f}_{2}}\right) \right). \label{GK}
\end{equation}

If we apply the result given by equation (\ref{apgz}) and consider the fact that the
spontaneous fission neutron source will behave as a Poissonian source, then we find
that (\citet{Bell65,BM2005})

\begin{equation}
G_{\rm S,\mathcal{Y}}\left( s, t_{\rm f}\right)= \exp \left( \int_{-\infty }^{t_{\rm f}}
\mathrm{d}t\int \mathrm{d}\mathbf{r}\lambda _{\rm sf}N\left( t\right) \rho _{\rm
sf}\left( \mathbf{r}\right) \left[s -1\right] \right) , \label{pgfGSY}
\end{equation}

\begin{eqnarray}
G_{\rm S}^{\rm sf}\left( s_{1},s_{2}|d_{1}\left( t_{{\rm
f}_{1}}\right),d_{2}\left(t_{{\rm f}_{2}}\right) \right) = G_{\rm S,\mathcal{Y}}\left( G_{\rm S,\mathcal{X}}\left(
s_{1},s_{2},\mathbf{r},t|d_{1}\left( t_{{\rm f}_{1}}\right),d_{2}\left(t_{{\rm
f}_{2}}\right) \right) , t_{\rm f}\right), \label{pgfbell}
\end{eqnarray} where we have supposed the source, which is given by the product of its time
dependent activity, $\lambda _{\rm sf}N\left( t\right) $, and the shape probability
distribution function $\rho _{\rm sf}\left( \mathbf{r}\right) $, to be introduced in the
remote past. In addition, the relationship between the spontaneous disintegration source
probability generating function and the kernel probability generating function is given
by (\citet{MPV87,MV87})

\begin{eqnarray}
G_{\rm S,\mathcal{X}}& &\left(s_{1},s_{2},\mathbf{r},t|d_{1}\left( t_{{\rm
f}_{1}}\right),d_{2}\left(t_{{\rm f}_{2}}\right) \right) \nonumber \\ & &  =\sum_{j=0}^{I_{\rm
sf}}\varepsilon _{j}^{\rm sf}\left[ \int \mathrm{d} v\int \mathrm{d}\mathbf{\Omega
}\frac{\chi _{\rm S}\left( v\right) }{4\pi } G_{\rm K}\left(s_{1},s_{2}, \vartheta
,t|d_{1}\left( t_{{\rm f}_{1}}\right),d_{2}\left(t_{{\rm f}_{2}}\right) \right) \right]
^{j}, \label{pgfdis}
\end{eqnarray} $\varepsilon_{j}^{\rm sf}$ being the probability of emission of $j$ neutrons
after a spontaneous fission within the fuel material, $I_{\rm sf}$ the maximum number of
spontaneous fission neutrons emitted after a source disintegration, and $\chi_{\rm S}(v)$
their corresponding spectrum.

On the other hand, in order to obtain the relationship applicable for the
proton-beam-driven spallation neutron source we can proceed in the following way:
firstly, let us consider the expression for the joint probability to register
$\mathcal{M}_{i}=m_{i}$, $i=1,2$, neutron counts in each neutron detector during the
corresponding detector time interval $\left( t_{{\rm f}_{i}}-\tau_{{\rm c}_{i}},t_{{\rm
f}_{i}}\right) $ following the injection of one single proton (1p) belonging to one
proton source pulse injected at a random time $\xi \in \left( 0,T\right) $, $T$ being the
pulsed proton source period, i.e., \citet{BM2005},

\begin{eqnarray}
P_{m_{1}m_{2}}^{\rm 1p}& &\left( \xi |d_{1}\left( t_{{\rm f}_{1}}\right),d_{2}\left(t_{{\rm
f}_{2}}\right) \right) \nonumber \\ & &= \int \mathrm{d}\mathbf{r}\rho _{\rm sp}\left(
\mathbf{r}\right) \sum_{j=0}^{I_{\rm sp}}\varepsilon _{j}^{\rm
sp}\sum_{m_{1}^{(1)},m_{2}^{(1)}=0}^{m_{1},m_{2}}\cdots
\sum_{m_{1}^{(j)},m_{2}^{(j)}=0}^{m_{1},m_{2}}\prod_{i=1}^{j}\int \mathrm{d}v_{i}\int
\mathrm{d}\mathbf{\Omega }_{i} \nonumber \\ & &\times f_{\rm sp}\left(
v_{i},\mathbf{\Omega }_{i}\right) K_{m_{1}^{(i)}m_{2}^{(i)}}\left(
\mathbf{r},v_{i},\mathbf{\Omega }_{i},\xi |d_{1}\left( t_{{\rm
f}_{1}}\right),d_{2}\left(t_{{\rm f}_{2}}\right) \right) , \label{poneproton}
\end{eqnarray} restricted by the constraints $m_{i}^{(1)}+m_{i}^{(2)}+\ldots m_{i}^{(j)}=m_{i}$,
$i=1,2$, where $\varepsilon _{j}^{\rm sp}$ is the probability of emission of $j$ neutrons
after a spallation interaction within the target material, $I_{\rm sp}$ the maximum
number of neutrons emitted in each spallation interaction, $\rho_{\rm sp}\left(
\mathbf{r} \right)$ is the spatial distribution function for neutrons born after a
spallation interaction within the target material, and $f_{\rm sp}\left(
v_{i},\mathbf{\Omega }_{i}\right) $ the spectral and angular probability distribution
function corresponding to these spallation neutrons. If we multiply the previous
expression by $s_{1}^{m_{1}}s_{2}^{m_{2}}$ and sum up from $m_{1},m_{2}=0$ to $\infty$,
we get the relationship between the spallation source probability generating function and
the kernel one for one proton randomly injected,

\begin{eqnarray}
G_{\rm 1p} &&\left( s_{1},s_{2},\xi |d_{1}\left( t_{{\rm f}_{1}}\right),d_{2}\left(t_{{\rm
f}_{2}}\right) \right)  \nonumber \\ && =
\int \mathrm{d}\mathbf{r}%
\rho _{\rm sp}\left( \mathbf{r}\right) \sum_{j=0}^{I_{\rm sp}}\varepsilon _{j}^{\rm
sp}\left[T_{\rm sp}G_{\rm K}\left(s_{1},s_{2}, \vartheta ,\xi |d_{1}\left( t_{{\rm
f}_{1}}\right),d_{2}\left(t_{{\rm f}_{2}}\right) \right) \right] ^{j}.
\label{pgfoneproton}
\end{eqnarray} where we have used the spallation operator $
T_{\rm sp}\circ =\int \mathrm{d}v_{i}\int \mathrm{d}\mathbf{\Omega } _{i}f_{\rm sp}\left(
v_{i},\mathbf{\Omega }_{i}\right) \circ$. With this expression and taking into account
the relationships expressed at the beginning of this Section, and admitting that the
number of detector counts after the introduction of each proton belonging to the same
source pulse can be considered as independent discrete random variables, we can derive
the relationship between the spallation source and the kernel probability generating
functions for the injection of $k$ protons in one proton pulse (pp) at time $\xi $,

\begin{eqnarray}
G_{{\rm pp},k}& &\left( s_{1},s_{2},\xi |d_{1}\left( t_{{\rm
f}_{1}}\right),d_{2}\left(t_{{\rm f}_{2}}\right) \right) \nonumber \\ & & =\left[ \int \mathrm{d}
\mathbf{r}\rho _{\rm sp}\left( \mathbf{r}\right) \sum_{j=0}^{I_{\rm sp}}\varepsilon
_{j}^{\rm sp}\left[ T_{\rm sp}G_{\rm K}\left(s_{1},s_{2}, \vartheta ,\xi |d_{1}\left(
t_{{\rm f}_{1}}\right),d_{2}\left(t_{{\rm f}_{2}}\right) \right) \right] ^{j}\right]
^{k}. \label{pgfonepulse}
\end{eqnarray} But, in general, the number of protons injected per accelerator pulse
might be considered to be a discrete random variable, thus, the correct expression for
the relationship between the spallation source and the kernel pgfs for the injection of
one proton pulse can be recast as

\begin{eqnarray}
G_{\rm pp}\left( s_{1},s_{2},\xi |d_{1}\left( t_{{\rm f}_{1}}\right),d_{2}\left(t_{{\rm
f}_{2}}\right) \right) = \sum_{k=0}^{I_{\rm pp}} \varepsilon_{k}^{\rm pp}  G_{{\rm
pp},k}\left( s_{1},s_{2},\xi |d_{1}\left( t_{{\rm f}_{1}}\right),d_{2}\left(t_{{\rm
f}_{2}}\right) \right), \label{pgfonepulse2}
\end{eqnarray} where $\varepsilon_{k}^{\rm pp}$ is the probability for the accelerator to
inject $k$ protons per proton pulse and $I_{\rm pp}$ is the maximum number of protons
that can be introduced in the system per proton pulse.

Considering this expression, for a periodic pulsed proton source such as that one given
by equation (\ref{spsource}), we will have

\begin{eqnarray}
G_{\rm S}^{\rm sp}\left( s_{1},s_{2},\xi |d_{1}\left( t_{{\rm
f}_{1}}\right),d_{2}\left(t_{{\rm f}_{2}}\right) \right) &=& \prod_{m=-\infty }^{\infty }
G_{\rm pp}\left( s_{1},s_{2} ,\xi +m T |d_{1}\left( t_{{\rm f}_{1}}\right),d_{2}\left(t_{{\rm
f}_{2}}\right) \right). \label{pgfspalxi}
\end{eqnarray}

In general, neutron detectors will register counts coming from both, spontaneous fission
and spallation, sources which can be treated as mutually independent random
variables (\citet{BM2005}), thus the probability generating function governing both
processes will be given by the product of equations (\ref{pgfbell}) and
(\ref{pgfspalxi}), that is,

\begin{eqnarray}
G_{\rm S} && \left( s_{1},s_{2},\xi |d_{1}\left( t_{{\rm f}_{1}}\right),d_{2}\left(t_{{\rm
f}_{2}}\right) \right) \nonumber \\ &&  = G_{\rm S}^{\rm sf}\left( s_{1},s_{2} |d_{1}\left( t_{{\rm
f}_{1}}\right),d_{2}\left(t_{{\rm f}_{2}}\right) \right) \times G_{\rm S}^{\rm sp}\left(
s_{1},s_{2},\xi |d_{1}\left( t_{{\rm f}_{1}}\right),d_{2}\left(t_{{\rm f}_{2}}\right)
\right) . \label{pgfxi}
\end{eqnarray}

This last equation expresses the relationship between the source probability generating
function and the kernel probability generating function when we consider the effect of
the intrinsic spontaneous fission source and the periodic pulsed spallation neutron
source, \citet{BM2005}. In particular, for the deterministic pulsed method we just need to
choose the elapsed time $\xi$ equal to zero, whereas, in order to apply the stochastic
pulsing method we will calculate the expected value of (\ref{pgfxi}), $\xi $ being
uniformly sampled between zero and the proton pulse period, $T$, i.e.,

\begin{eqnarray}
G_{\rm S} \left( s_{1},s_{2}|d_{1}\left( t_{{\rm f}_{1}}\right),d_{2}\left(t_{{\rm
f}_{2}}\right) \right) &=&\left\langle G_{\rm S}\left( s_{1},s_{2},\xi |d_{1}\left( t_{{\rm
f}_{1}}\right),d_{2}\left(t_{{\rm f}_{2}}\right) \right) \right\rangle _{\xi
} \nonumber \\ & =&\int_{0}^{T}\frac{\mathrm{d}\xi}{T} G_{\rm S}\left( s_{1},s_{2},\xi |d_{1}\left(
t_{{\rm f}_{1}}\right),d_{2}\left(t_{{\rm f}_{2}}\right) \right) . \label{pgf}
\end{eqnarray}


\section{The Boltzmann neutron transport equation for counting problems from
the stochastic neutron transport theory}

In this Section we derive the integro-differential equation governing the kernel
probability generating function $G_{\rm K}\left(s_{1},s_{2}, \vartheta ,\xi |d_{1}\left(
t_{{\rm f}_{1}}\right),d_{2}\left(t_{{\rm f}_{2}}\right) \right) $. First of all, we need
to obtain an expression for the probability function $K_{z_{1}z_{2}}\left( \vartheta
,t|d_{1}\left( t_{{\rm f}_{1}}\right),d_{2}\left(t_{{\rm f}_{2}}\right) \right) $ for
both neutron detectors. It can be done using a probability balance of mutually exclusive
events (\citet{MPV87,Munoz2000}). Then we shall multiply the probability balance equation of
$K_{z_{1}z_{2}}\left( \vartheta ,t|d_{1}\left( t_{{\rm f}_{1}}\right),d_{2}\left(t_{{\rm
f}_{2}}\right) \right) $ by the factor $s_{1}^{z_{1}}s_{2}^{z_{2}}$ and then sum up from
$z_{1},z_{2}=0$ to $\infty $. Next, we need to apply the known P\'{a}l's
methodology (see \citet{Pal58,Bell65,MPV87}) to obtain the non-linear transport
integro-differential equation satisfied by the kernel probability generating
function, \citet{MPV87}:

\begin{eqnarray}
\mathcal{H}G_{\rm K} && \left(s_{1},s_{2}, \vartheta ,t |d_{1}\left( t_{{\rm
f}_{1}}\right),d_{2}\left(t_{{\rm f}_{2}}\right) \right) \nonumber \\ && = \mathcal{S} \left( G_{\rm
K}\left(s_{1},s_{2}, \mathbf{r},v^{\prime },\mathbf{\Omega }^{\prime } ,t |d_{1}\left(
t_{{\rm f}_{1}}\right),d_{2}\left(t_{{\rm f}_{2}}\right) \right)\right), \label{HGKS}
\end{eqnarray} where we have defined the general time-dependent transport operator

\begin{equation}
\mathcal{H}=-\left( \frac{1}{v}\frac{\partial }{\partial t}+\mathbf{%
\Omega }\cdot \mathbf{\nabla }-\Sigma _{\rm t}\left( \mathbf{r},v\right) \right) ,
\label{Hoperator}
\end{equation} and the non-linear kernel pgf source operator

\begin{eqnarray}
\mathcal{S} \left( \circ\right) &=& \Sigma_{\rm t}\left( \mathbf{r},v \right)
\left\{ C_{\rm c}^{\left( 0,0\right) }\left( \mathbf{r%
},v,t\right) +s_{1} C_{\rm c}^{\left( 1,0\right) }\left( \mathbf{r},v,t\right) +
s_{2}C_{\rm c}^{\left( 0,1\right) }\left( \mathbf{r},v,t\right)\right. \nonumber \\ & & +
\left( C_{\rm s}^{\left( 0,0\right) }\left( \mathbf{r},v,t\right) +s_{1} C_{\rm
s}^{\left( 1,0\right) }\left( \mathbf{r},v,t\right) +s_{2} C_{\rm s}^{\left( 0,1\right)
}\left( \mathbf{r},v,t\right) \right)  \nonumber \\ &\times& \int \mathrm{d}v^{\prime
}\int \mathrm{d}\mathbf{\Omega }^{\prime}  f_{\rm s}\left(\mathbf{r}, v,\mathbf{\Omega
}|v^{\prime },\mathbf{\Omega }^{\prime }\right) \circ \nonumber \\ & & + \sum_{j=0}^{I}
\left( C_{j}^{\left( 0,0\right) }\left( \mathbf{r},v,t\right) +s_{1} C_{j}^{\left(
1,0\right) }\left( \mathbf{r},v,t\right) +s_{2} C_{j}^{\left( 0,1\right) }\left(
\mathbf{r},v,t\right) \right) \nonumber \\
&\times& \left. \left[ \int \mathrm{d}v^{\prime }\int \mathrm{d}\mathbf{%
\Omega }^{\prime }\frac{\chi \left(\mathbf{r}, v^{\prime }\right) }{4\pi }\circ \right]
^{j} \right\}, \label{SGK}
\end{eqnarray} where

\[
f_{\rm s}\left(\mathbf{r}, v,\mathbf{\Omega }|v^{\prime },\mathbf{\Omega }^{\prime
}\right) = \left\{ \begin{array}{ll} f_{\rm s}\left( v,\mathbf{\Omega }|v^{\prime
},\mathbf{\Omega }^{\prime }\right) & \mbox{for } \mathbf{r}\notin V_{\rm D_{1}},V_{\rm
D_{2}}, \\ f_{\rm s}^{{\rm D}_{i}} \left( v,\mathbf{\Omega }|v^{\prime },\mathbf{\Omega
}^{\prime }\right) & \mbox{for } \mathbf{r} \in V_{{\rm D}_{i}},
\end{array} \right.
\]%
represents the probability distribution function for a neutron to exit with velocity and
direction within $\left( v^{\prime },v^{\prime }+ \mathrm{d}v^{\prime }\right) $ and
$\left( \mathbf{\Omega }^{\prime },\mathbf{\Omega }^{\prime }+\mathrm{d}\mathbf{\Omega
}^{\prime }\right) $, respectively, after a scattering event (the superscript ${\rm
D}_{i}$ applies for the $i$-th detector volume) with an incident neutron with velocity
$v$ and direction $\mathbf{\Omega }$, whereas the spectrum of neutrons emitted following
a fission event is given by

\[
\chi \left(\mathbf{r}, v\right) = \left\{ \begin{array}{ll} \chi \left( v\right) &
\mbox{for } \mathbf{r}\notin V_{\rm D_{1}},V_{\rm D_{2}}, \\ \chi_{{\rm D}_{i}} \left(
v\right) & \mbox{for } \mathbf{r} \in V_{{\rm D}_{i}},
\end{array} \right.
\]%
that is, as before, for the system volume no subscript is used, while to refer to
fissions occurring within either detector volume we will add the subscript ${\rm D}_{i}$.
$I$ is the maximum number of neutrons produced after a fission event.

In addition, in expression (\ref{SGK}) we must specify the C-probabilities:

\begin{equation}
C_{\rm c}^{\left( 0,0\right) }\left( \mathbf{r},v,t\right) =\left\{
\begin{array}{cc}
\frac{\Sigma _{\rm c}}{\Sigma _{\rm t}} & \mbox{if }\mathbf{r}\notin V_{\rm D_{1}},V_{\rm
D_{2}}, \\ \frac{\Sigma _{\rm c}^{{\rm D}_{1}}}{\Sigma _{\rm t}^{{\rm D}_{1}}}
\left(1-\eta_{\rm c}^{{\rm D}_{1}} \Delta\left(d_{1}\left(t_{{\rm
f}_{1}}\right)\right)\right) & \mbox{if }\mathbf{r}\in
V_{\rm D_{1}}, \\ %
\frac{\Sigma _{\rm c}^{\rm D_{2}}}{\Sigma _{\rm t}^{\rm D_{2}}}\left(1-\eta_{\rm c}^{\rm
D_{2} }\Delta\left(d_{2}\left(t_{{\rm f}_{2}}\right)\right)\right) & \mbox{if
}\mathbf{r}\in V_{\rm D_{2}},
\end{array}%
\right.  \label{c000}
\end{equation} is the probability to have zero detector counts following a capture event within
the nuclear system ($V_{\rm SYS}$) or within one of the detectors ($V_{{\rm D}_{i}}$,
$i=1,2$) after a given neutron interaction at position $\mathbf{r}$ and time $t$.
$\Sigma_{\rm t}$ ($\Sigma_{\rm t}^{{\rm D}_{i}}$) denotes the neutron total macroscopic
cross section for the system ($i$-th detector) volume, $\Sigma_{\rm c}$ ($\Sigma_{\rm
c}^{{\rm D}_{i}}$) is the neutron capture macroscopic cross section, and $\eta_{\rm c
}^{{\rm D}_{i}}$ accounts for the $i$-th detector capture efficiency. $\Delta
\left(d_{i}\left(t_{{\rm f}_{i}}\right)\right) =\left( H\left( t-\left( t_{{\rm
f}_{i}}-\tau _{{\rm c}_{i}}\right) \right) -H\left( t-t_{{\rm f}_{i}}\right) \right) $ is
the time window for the $i$-th detector, $H\left( t\right) $ being the characteristic or
Heaviside function. Similarly, for one neutron count after a capture event in the first
detector, we have

\begin{equation}
C_{\rm c}^{\left( 1,0\right) }\left( \mathbf{r},v,t\right) =\left\{
\begin{array}{cc}
0 & \mbox{if }\mathbf{r}\notin V_{\rm D_{1}}, \\ \eta_{\rm c}^{\rm D_{1}}
 \frac{\Sigma _{\rm c}^{\rm D_{1}}}{\Sigma _{\rm t}^{\rm D_{1}}}
\Delta\left(d_{1}\left(t_{{\rm f}_{1}}\right)\right) & \mbox{if }\mathbf{r}\in V_{\rm
D_{1}}.
\end{array}%
\right. \label{c001}
\end{equation}

The expression corresponding to the case of one neutron count following a neutron capture
event in the second detector can be derived in an analogous way.

Next, in equation (\ref{SGK}) we must also specify the probability to have zero
counts in both detectors after a neutron scattering event at position $\mathbf{r}$ and
time $t$, $v$ being the incident neutron velocity, i.e.,

\begin{equation}
C_{\rm s}^{\left( 0,0\right) }\left( \mathbf{r},v,t\right) =\left\{
\begin{array}{cc}
\frac{\Sigma _{\rm s}}{\Sigma _{\rm t}} & \mbox{if }\mathbf{r}\notin V_{\rm D_{1}},V_{\rm
D_{2}} \\ \frac{\Sigma _{\rm s}^{\rm D_{1}}}{\Sigma _{\rm t}^{\rm D_{1}}}\left( 1-\eta
_{\rm s}^{\rm D_{1}}\Delta\left(d_{1}\left(t_{{\rm f}_{1}}\right)\right) \right) %
& \mbox{if }\mathbf{r}\in V_{\rm D_{1}},%
\\ \frac{\Sigma _{\rm s}^{\rm D_{2}}}{\Sigma _{\rm t}^{\rm D_{2}}}
\left( 1-\eta _{\rm s}^
{\rm D_{2}}\Delta\left(d_{2}\left(t_{{\rm f}_{2}}\right)\right) \right) %
& \mbox{if }\mathbf{r}\in V_{\rm D_{2}},%
\end{array}
\right.  \label{c100}
\end{equation}%
where $\Sigma_{\rm s}$ ($\Sigma_{\rm s}^{{\rm D}_{i}}$) denotes the neutron scattering
macroscopic cross section for the system ($i$-th detector) volume and $\eta_{\rm s
}^{{\rm D}_{i}}$ is the $i$-th detector scattering efficiency.

In case of one detector count, for instance, in the first detector:

\begin{equation}
C_{\rm s}^{\left( 1,0\right) }\left( \mathbf{r},v,t\right) =\left\{
\begin{array}{cc}
0 & \mbox{if }\mathbf{r}\notin V_{\rm D_{1}}, \\ \eta _{\rm s}^{\rm D_{1}}  \frac{\Sigma
_{\rm s}^{\rm D_{1}}}{\Sigma _{\rm t}^{\rm D_{1}}}\Delta \left(d_{1}\left(t_{{\rm
f}_{1}}\right)\right) & \mbox{if }\mathbf{r}\in V_{\rm D_{1}}.%
\end{array}%
\right. \label{c110}
\end{equation}%

Finally, in equation (\ref{SGK}) the probability of occurrence of a fission event with
emission of $j\geq 0$ neutrons leading to zero detector counts following a neutron
interaction at position $\mathbf{r}$ and time $t$ is given by

\begin{equation}
C_{j}^{\left( 0,0\right) }\left( \mathbf{r},v,t\right) =\left\{
\begin{array}{cc}
\varepsilon _{j}\frac{\Sigma _{\rm f}}{\Sigma _{\rm t}} & \mbox{if }\mathbf{r}\notin
V_{\rm D_{1}},V_{\rm D_{2}}, \\ \varepsilon _{j}^{\rm D_{1}}\frac{\Sigma _{\rm f}^{\rm
D_{1}}}{\Sigma _{\rm t}^{\rm D_{1}}}\left( 1-\eta _{\rm f}^{\rm D_{1}}\Delta
\left(d_{1}\left(t_{{\rm f}_{1}}\right)\right) \right) &
\mbox{if }\mathbf{r}\in V_{\rm D_{1}},%
\\ \varepsilon _{j}^{\rm D_{2}}\frac{\Sigma _{\rm f}^{\rm
D_{2}}}{\Sigma _{\rm t}^{\rm D_{2}}}\left( 1-\eta _{\rm f}^{\rm D_{2}}\Delta
\left(d_{2}\left(t_{{\rm f}_{2}}\right)\right) \right) &
\mbox{if }\mathbf{r}\in V_{\rm D_{2}},%
\end{array}%
\right.  \label{cj00}
\end{equation}%
where $\varepsilon_{j}$ ($\varepsilon _{j}^{{\rm D}_{i}}$) accounts for the probability
to emit $j$ neutrons after a fission event within the system ($i$-th detector) volume,
$\Sigma_{\rm f}$ ($\Sigma_{\rm f}^{{\rm D}_{i}}$) denotes the neutron fission macroscopic
cross section for the system ($i$-th detector) volume and $\eta_{\rm f}^{{\rm D}_{i}}$ is
the $i$-th detector fission efficiency. Whereas, if we consider, e.g., one detector count
registered by the first detector,

\begin{equation}
C_{j}^{\left( 1,0\right) }\left( \mathbf{r},v,t\right) =\left\{
\begin{array}{cc}
0 & \mbox{if }\mathbf{r}\notin V_{\rm D_{1}}, \\ \eta _{\rm f}^{\rm D_{1}}\varepsilon
_{j}^{\rm D_{1}}\frac{\Sigma _{\rm f}^{\rm D_{1}}}{\Sigma _{\rm t}^{\rm D_{1}}} \Delta
\left( d_{1}\left(t_{{\rm f}_{1}}\right)\right) &
\mbox{if }\mathbf{r}\in V_{\rm D_{1}}.%
\end{array}%
\right.   \label{cj10}
\end{equation}

Expression (\ref{HGKS}) must fulfil the final condition $G_{\rm K}\left(
s_{1},s_{2},\vartheta ,t|d_{1}\left( t_{{\rm f}_{1}}\right),d_{2}\left(t_{{\rm
f}_{2}}\right) \right) =1$ for $t>t_{\rm f}=\max\left\{ t_{{\rm f}_{1}},t_{{\rm f}_{2}}
\right\}$, due to the causality principle, and the boundary condition $G_{\rm K}\left(
s_{1},s_{2},\mathbf{r}_{\rm B},v,\mathbf{\Omega },t|d_{1}\left( t_{{\rm
f}_{1}}\right),d_{2}\left(t_{{\rm f}_{2}}\right) \right) =1$ for $\mathbf{n}\cdot
\mathbf{\Omega }>0$, i.e., for neutrons injected outwardly at a convex
boundary (\citet{Bell65,MPV87}).

Further, according to Bartlett's procedure, \citet{Bartlett}, we can derive the first
factorial moment of the number of detector counts per single neutron injected in the
system at the phase-space point $\vartheta $ and at time $t$ as

\begin{equation}
\bar{z}_{i}\left( \vartheta ,t|d\left( t_{\rm f}\right) \right) =\left. \frac{%
\partial }{\partial s_{i}}G_{\rm K}\left( s_{1},s_{2},\vartheta ,t|d_{i}\left( t_{{\rm f}_{i}}
\right)\right) \right\vert _{s_{1},s_{2}=1},  \label{bartlettzi}
\end{equation}%
whereas the cross second factorial moment of the number of detector counts per single
neutron injected can be defined as

\begin{equation}
\overline{z_{1}z_{2}}\left( \vartheta ,t|d\left( t_{\rm f}\right) \right) =\left.
\frac{\partial ^{2}}{\partial s_{1}\partial s_{2}}G_{\rm K}\left( s_{1},s_{2},\vartheta
,t|d_{1}\left( t_{\rm f_{1}}\right),d_{2}\left( t_{\rm f_{2}}\right) \right) \right\vert
_{s_{1},s_{2}=1}. \label{bartlettz1z2}
\end{equation}

Thus, applying the operator $\left. \partial /\partial s_{i}\right\vert _{s_{1},s_{2}=1}$
to the expression corresponding to the transport integro-differential equation satisfied
by $G_{\rm K}\left( s_{1},s_{2},\vartheta ,t|d_{1}\left( t_{\rm f_{1}}\right),d_{2}\left(
t_{\rm f_{2}}\right) \right)$, equation (\ref{HGKS}), we get%

\begin{equation}
\left( -\frac{1}{v}\frac{\partial }{\partial t} - L^{+}\right) \bar{z}_{i}%
=S_{{\rm D}_{i}}^{+},  \label{transportzi+}
\end{equation} where the time-independent adjoint transport operator $L^{+}$, \citet{MPV87}, and

\begin{eqnarray}
S_{{\rm D}_{i}}^{+} \left( \vartheta, t \right) = S_{{\rm D}_{i}}^{+} \left( \vartheta
\right) \times S_{{\rm D}_{i}}^{+}\left( t\right) & = &\left[\eta _{\rm c}^{{\rm
D}_{i}}\Sigma _{\rm c}^{{\rm D}_{i}}+ \eta _{\rm s}^{{\rm D}_{i}}\Sigma _{\rm s}^{{\rm
D}_{i}} +\eta _{\rm f}^{{\rm D}_{i}}\Sigma _{\rm f}^{{\rm D}_{i}}\right] \nonumber \\
&\times& \left[ H\left( t-\left( t_{{\rm f}_{i}}-\tau _{{\rm c}_{i}}\right) \right)
-H\left( t-t_{{\rm f}_{i}}\right) \right]. \label{SDi+}
\end{eqnarray} This magnitude will be non-zero only for $\mathbf{r}$ $\in$ $V_{{\rm D}_{i}}$ and
$t$ $\in$ $\left( t_{{\rm f}_{i}}-\tau_{{\rm c}_{i}}, t_{{\rm f}_{i}} \right]$.

The solution $\bar{z}_{i}$ of equation (\ref{transportzi+}) must fulfil the boundary
condition $\bar{z}_{i}=0$ for $\mathbf{n \cdot \Omega}>0$, on a convex boundary, and the
time-reversed causality condition, i.e., it must vanish at the end of the measurement
period $\tau _{{\rm c}_{i}}$. Indeed, equation (\ref{transportzi+}) reveals the nature of
$\bar{z}_{i}$ as an adjoint generalised Green's function driven by the adjoint importance
source $S_{{\rm D}_{i}}^{+}$, \citet{MPV87}. Consequently, for the forward transport problem
we shall write

\begin{equation}
\left( \frac{1}{v}\frac{\partial }{\partial t}- L\right) \phi =S_{1}, \label{transportn1}
\end{equation} where $L$ is the time-independent direct transport operator, \citet{Bell79}, and

\begin{eqnarray}
S_{1}\left( \vartheta, t \right) &=& S_{1}^{\rm sf}\left( \vartheta, t \right) +
S_{1}^{\rm sp}\left( \vartheta, t \right) \nonumber \\ &=& S_{1}^{\rm sf}\left( \vartheta
\right) \times S_{1}^{\rm sf}\left( t \right) + S_{1}^{\rm sp}\left( \vartheta \right)
\times S_{1}^{\rm sp}\left( t \right) , \label{S1}
\end{eqnarray} that is, the total neutron source can be expressed as the sum of the intrinsic
and the spallation neutron sources,

\begin{equation}
S_{1}^{\rm sf}\left( \vartheta \right) =\bar{\nu} _{\rm sf} N_{0} \rho_{\rm sf} \left(
\mathbf{r} \right) \frac{\chi_{\rm S} \left( v \right)}{4 \pi}, \label{S1sftheta}
\end{equation}

\begin{equation}
S_{1}^{\rm sf}\left( t\right) =\lambda _{\rm sf}N\left( t\right) /N_{0}\equiv
\lambda _{\rm sf},  \label{S1sft}
\end{equation} where we assume the initial number of nuclei corresponding to the
spontaneous disintegration neutron source to be constant in our time-scale,

\begin{equation}
S_{1}^{\rm sp}\left( \vartheta \right) = \bar{\nu}_{\rm pp} \bar{\nu}_{\rm sp} \rho_{\rm
sp} \left( \mathbf{r} \right)  f_{\rm sp} \left( v, \mathbf{\Omega} \right) ,
\label{S1sptheta}
\end{equation}

\begin{equation}
S_{1}^{\rm sp}\left( t\right) =\sum_{m=-\infty }^{\infty}\delta \left( t-\left( \xi
+mT\right)\right) , \label{S1spt}
\end{equation} with $\bar{\nu}_{\rm w}=\sum_{1}^{I_{\rm w}}j\varepsilon _{j}^{\rm w}$,
${\rm w=sf,sp,pp}$.

Now, the forward neutron flux satisfies the initial condition $\phi \left(t=-\infty
\right) =0$, and the boundary condition $\phi =0$ for $\mathbf{n \cdot \Omega}<0$, on a
convex boundary. A proper choice of the corresponding boundary and final conditions for
the adjoint function makes the associated bilinear concomittance to vanish, and, hence,
due to the commutation relation

\begin{equation}
\left\langle \bar{z}_{i}|S_{1}\right\rangle =\left\langle S_{{\rm D}_{i}}^{+}|\phi
\right\rangle , \label{comreln1}
\end{equation} where Dirac's notation for the inner product is used, and where both terms
account for the average number of detector counts during its counting interval: at the
left hand side we have the inner product of the neutron source strength, $S_{1}$,
(neutrons emitted at a given phase-space point and time) and the number of counts
gathered at detector $i$ per single neutron introduced at a given phase-space point and
time, $\bar{z}_{i}$; equivalently, the right hand side term expresses the inner product
of the effective macroscopic neutron detection cross section for the same detector,
$S_{{\rm D}_{i}}^{+}$, and the neutron flux, $\phi $, \citet{Munoz2000}.

Similarly, we can make use of (\ref{bartlettz1z2}) to find an expression for the cross second factorial moment of the number of detector counts per single neutron introduced:

\begin{equation}
\left( -\frac{1}{v}\frac{\partial }{\partial t} - L^{+}\right)
\overline{z_{1}z_{2}}=S_{\rm D_{1}D_{2}}^{+},
 \label{transportz1z2+}
\end{equation} where the importance source for the cross second factorial moment can be recast as

\begin{eqnarray}
&& \frac{S_{\rm D_{1}D_{2}}^{+}\left( \vartheta, t \right)}{\Sigma_{\rm
t}\left(\mathbf{r},v\right)} \nonumber \\ && = C_{\rm s}^{\left(1,0\right)}\left(
\mathbf{r},v,t\right) \int{\rm d}v^{\prime}\int{\rm d}\mathbf{\Omega}^{\prime}f_{\rm s
}\left(\mathbf{r},v,\mathbf{\Omega}|v^{\prime},\mathbf{\Omega}^{\prime}\right)
\bar{z}_{1}\left(\mathbf{r},v^{\prime},\mathbf{\Omega}^{\prime},t|d_{1}\left( t_{\rm
f_{1}} \right)\right) \nonumber \\ %
&& +C_{\rm s}^{\left(0,1\right)}\left( \mathbf{r},v,t\right) \int{\rm
d}v^{\prime}\int{\rm d}\mathbf{\Omega}^{\prime}f_{\rm s
}\left(\mathbf{r},v,\mathbf{\Omega}|v^{\prime},\mathbf{\Omega}^{\prime}\right)
\bar{z}_{2}\left(\mathbf{r},v^{\prime},\mathbf{\Omega}^{\prime},t|d_{2}\left( t_{\rm
f_{2}} \right)\right) \nonumber \\ %
&& +\sum_{j=1}^{I}jC_{j}^{\left(1,0\right)}\left( \mathbf{r},v,t\right) \int{\rm
d}v^{\prime}\int{\rm d}\mathbf{\Omega}^{\prime}\chi\left(\mathbf{r},v^{\prime}\right)/
4\pi \bar{z}_{1}\left(\mathbf{r},v^{\prime},\mathbf{\Omega}^{\prime},t|d_{1}\left( t_{\rm
f_{1}} \right)\right) \nonumber \\ %
&& +\sum_{j=1}^{I}jC_{j}^{\left(0,1\right)}\left( \mathbf{r},v,t\right) \int{\rm
d}v^{\prime}\int{\rm d}\mathbf{\Omega}^{\prime}\chi\left(\mathbf{r},v^{\prime}\right)/
4\pi \bar{z}_{2}\left(\mathbf{r},v^{\prime},\mathbf{\Omega}^{\prime},t|d_{2}\left( t_{\rm
f_{2}} \right)\right) \nonumber \\ %
&& +\sum_{j=2}^{I}j\left(j-1\right)
\sum_{r_{1},r_{2}=0, r_{1}+r_{2}=1}^{1}C_{j}^{\left(r_{1},r_{2}\right)}\left(
\mathbf{r},v,t\right) \int{\rm d}v^{\prime}\int{\rm
d}\mathbf{\Omega}^{\prime}\chi\left(\mathbf{r},v^{\prime}\right)/ 4\pi \nonumber \\ %
&& \times  \bar{z}_{1}\left(\mathbf{r},v^{\prime},\mathbf{\Omega}^{\prime},t|d_{1}\left(
t_{\rm f_{1}} \right)\right)
\bar{z}_{2}\left(\mathbf{r},v^{\prime},\mathbf{\Omega}^{\prime},t|d_{2}\left( t_{\rm
f_{2}} \right)\right) . \label{SD1D2+}
\end{eqnarray}

The cross second factorial moment $\overline{z_{1}z_{2}}$ must satisfy the same
time-reversed and boundary conditions as $\bar{z}_{i}$, but for $t_{\rm f}=\max\{t_{\rm
f_{1}},t_{\rm f_{2}}\}$. Hence, it can be viewed as a generalised adjoint function, now
driven by the adjoint source, $S_{\rm D_{1}D_{2}}^{+}$, that is, the product of the detector
cross sections and the spectral and angular weighted neutron importances. Again, the
commutation relation leads to the identity

\begin{equation}
\left\langle \overline{z_{1}z_{2}}|S_{1}\right\rangle =\left\langle S_{\rm
D_{1}D_{2}}^{+}|\phi \right\rangle .  \label{comreln2}
\end{equation}

In order to express the adjoint problem in terms of instantaneous detector counting
rates, we shall divide adjoint transport equations (\ref{transportzi+}) and
(\ref{transportz1z2+}) by $\tau_{{\rm c}_{i}}$ and $\tau_{{\rm c}_{1}}\tau_{{\rm c}_{2}}$
and then calculate the limits $\lim_{\tau_{{\rm c}_{i}}\downarrow 0}$ and
$\lim_{\tau_{{\rm c}_{1}},\tau_{{\rm c}_{2}}\downarrow0}$, respectively. As a
consequence, we shall write

\begin{equation}
\left( -\frac{1}{v}\frac{\partial }{\partial t} - L^{+}\right) \dot{\bar{z}}_{i}%
=\dot{S}_{{\rm D}_{i}}^{+},  \label{transportzi+dot}
\end{equation}

\begin{equation}
\left( -\frac{1}{v}\frac{\partial }{\partial t} - L^{+}\right)
\ddot{\overline{z_{1}z_{2}}}=\ddot{S}_{\rm D_{1}D_{2}}^{+},
 \label{transportz1z2+dot}
\end{equation} where, by definition,

\begin{equation}
\dot{\bar{z}}_{i}\left(\vartheta,t-t_{{\rm f}_{i}}\right)=\lim_{\tau_{{\rm
c}_{i}}\downarrow0} \frac{\bar{z}_{i}\left(\vartheta,t|d_{i}\left(t_{{\rm
f}_{i}}\right)\right)}{\tau_{{\rm c}_{i}}}, \label{zidot}
\end{equation}

\begin{equation}
\dot{S}_{{\rm D}_{i}}^{+}\left(\vartheta,t-t_{{\rm f}_{i}}\right)=\lim_{\tau_{{\rm
c}_{i}}\downarrow0}\frac{S_{{\rm D}_{i}}^{+}\left(\vartheta,t|d_{i}\left(t_{{\rm
f}_{i}}\right)\right)}{\tau_{{\rm c}_{i}}}, \label{SDi+dot}
\end{equation} since, taking into account equations (\ref{transportzi+dot}) and (\ref{SDi+dot}),
$\dot{\bar{z}}_{i}\left(\vartheta,t-t_{{\rm f}_{i}}\right)$ can be regarded as a
displacement kernel, \citet{MPV87},

\begin{equation}
\ddot{\overline{z_{1}z_{2}}}\left(\vartheta,t-t_{{\rm f}_{1}},t-t_{{\rm
f}_{2}}\right)=\lim_{\tau_{{\rm c}_{1}},\tau_{{\rm
c}_{2}}\downarrow0}\frac{\overline{z_{1}z_{2}}\left(\vartheta,t|d_{1}\left(t_{{\rm
f}_{1}}\right),d_{2}\left(t_{{\rm f}_{2}}\right)\right)}{\tau_{{\rm c}_{1}}\tau_{{\rm
c}_{2}}}, \label{z1z2dot}
\end{equation}

\begin{equation}
\ddot{S}_{\rm D_{1}D_{2}}^{+}\left(\vartheta,t-t_{{\rm f}_{1}},t-t_{{\rm
f}_{2}}\right)=\lim_{\tau_{{\rm c}_{1}},\tau_{{\rm c}_{2}}\downarrow0} \frac{S_{\rm
D_{1}D_{2}}^{+}\left(\vartheta,t|d_{1}\left(t_{{\rm f}_{1}}\right),d_{2}\left(t_{{\rm
f}_{2}}\right)\right)}{\tau_{{\rm c}_{1}}\tau_{{\rm c}_{2}}}. \label{SD1D2+dot}
\end{equation}

Next, we look for a solution to the Boltzmann neutron transport equation
(\ref{transportn1}) for the direct flux, $\phi \left( \vartheta ,t\right) $, satisfying
the $\alpha$-modes expansion:

\begin{equation}
\phi \left( \vartheta ,t\right)=\phi^{\rm sf} \left( \vartheta ,t\right)+\phi^{\rm sp}
\left( \vartheta ,t\right) =\sum_{j}\varphi _{j}\left( \vartheta \right) \zeta _{j}\left(
t\right) , \label{fluxexp}
\end{equation} where we assume the eigenfunctions $\varphi _{j}\left( \vartheta \right) $
to form a complete basis in the corresponding Hilbert space (\citet{Bell79,Carta99}). These
must obey the $\alpha $-eigenvalue equation

\begin{equation}
L\varphi _{j}\left( \vartheta \right) = \frac{\alpha _{j}}{v}\varphi
_{j}\left( \vartheta \right) .  \label{alphadirect}
\end{equation} Similarly, for the adjoint flux instantaneous rate we will have

\begin{equation}
\dot{\bar{z}}_{i}\left( \vartheta ,t-t_{{\rm f}_{i}} \right) =\sum_{j}\varphi _{{\rm
D}_{i}j}^{+}\left( \vartheta \right) \zeta_{{\rm D}_{i}j}^{+}\left( t-t_{{\rm
f}_{i}}\right) , \label{zi+exp}
\end{equation} and the $\alpha $-eigenvalue equation

\begin{equation}
L^{+}\varphi _{{\rm D}_{i}j}^{+}\left( \vartheta \right) = \frac{\alpha _{j}}{v}\varphi
_{{\rm D}_{i}j}^{+}\left( \vartheta \right) .  \label{alphaadjoint}
\end{equation} Beneath this ansatz, it is obvious that we must put $\varphi _{{\rm D}_{1}
j} ^{+}\left( \vartheta \right) =\varphi _{{\rm D}_{2} j}^{+}\left( \vartheta \right)
\equiv \varphi _{j} ^{+}\left( \vartheta \right)$. The adjoint and forward eigenfunctions
satisfy the biorthogonal relation, \citet{Bell79}, i.e.,

\begin{equation}
\left( \frac{1}{v}\varphi _{n}^{+},\varphi _{m}\right) =\delta _{nm}\left(
\frac{1}{v}\varphi _{n}^{+},\varphi _{n}\right) ,  \label{biortho}
\end{equation} where the phase-space inner product is defined by $\left( a,b\right) =\int
\mathrm{d}\mathbf{r} \mathrm{d}v \mathrm{d}\mathbf{\Omega }a\left( \vartheta \right)
b\left( \vartheta \right) $. If we introduce the ansatz (\ref{fluxexp}) in equation
(\ref{transportn1}) and apply the Fourier transform operator to both sides of it, we
shall obtain the Fourier transform of the $j$-th flux instataneous rate time-dependent term, which, on
account of identities (\ref{alphadirect}), (\ref{alphaadjoint}), and (\ref{biortho}),
reads as

\begin{eqnarray}
\zeta _{j}\left( \omega\right) = \frac{1}{{\rm i}\omega-\alpha_{j}} \left[ 2\pi
\lambda_{\rm
sf }\frac{\left( S_{1}^{\rm sf},\varphi _{j}^{+} \right) }{\left( \frac{1}{v}%
\varphi _{j} ,\varphi _{j}^{+} \right) } \delta\left( \omega\right)  + \frac{\left(
S_{1}^{\rm sp} ,\varphi _{j}^{+} \right) }{\left( \frac{1}{v}\varphi _{j} ,\varphi
_{j}^{+} \right) } \sum_{m=-\infty}^{\infty} {\rm e}^{-{\rm
i}\omega\left(\xi+mT\right)}\right] , \label{zetj}
\end{eqnarray} with

\begin{equation}
\left( S_{1}^{\rm sf} ,\varphi _{
j}^{+}\right) =\int \mathrm{d}\mathbf{r}\int \mathrm{d}v\int \mathrm{d}%
\mathbf{\Omega }\bar{\nu}_{\rm sf}N_{0}\rho _{\rm sf}\left( \mathbf{r}\right) \frac{%
\chi_{\rm S} \left( v\right) }{4\pi }\varphi _{j}^{+}\left( \mathbf{r},v,\mathbf{%
\Omega }\right) ,  \label{inprodS1sf}
\end{equation}%

\begin{equation}
\left( S_{1}^{\rm sp} ,\varphi _{j}^{+} \right) =\int \mathrm{d}\mathbf{r}\int
\mathrm{d}v\int \mathrm{d}%
\mathbf{\Omega } \bar{\nu}_{\rm pp} \bar{\nu}_{\rm sp} \rho _{\rm sp}\left(
\mathbf{r}\right) f_{\rm sp}\left( v,\mathbf{\Omega }\right) \varphi _{j}^{+}
\left( \mathbf{r},v,%
\mathbf{\Omega }\right) ,  \label{inprodS1sp}
\end{equation}

In the same way, from (\ref{transportzi+dot}) we can deduce the expression corresponding
to the Fourier transform of the $j$-th adjoint flux instantaneous rate time-dependent
term for the $i$-th neutron detector:

\begin{equation}
\zeta _{j}^{+}\left( \omega\right) = - \frac{1}{{\rm i}\omega+\alpha_{j}}
\frac{\left( S_{{\rm D}_{i}}^{+} ,\varphi _{j} \right) }{\left( \frac{1}{v}\varphi
_{j}^{+} ,\varphi _{j} \right) } . \label{zeti+}
\end{equation} where $S_{{\rm D}_{i}}^{+}=S_{{\rm D}_{i}}^{+}\left(\vartheta\right)$ is
given by (\ref{SDi+}).


\section{Analytical expression for the cross power spectral density with pulsed
sources}\label{cpsdsec}

\subsection{The deterministic pulsing method}

In order to obtain the analytical expressions corresponding to the factorial moments of
the number of counts of both neutron detectors we can apply again Bartlett's
procedure (\citet{Bartlett}) to the expression corresponding to the source probability
generating function. As we have outlined previously, we can do it taking into account two
different situations: in the first case, we can calculate factorial moments corresponding
to the deterministic pulsing method.

We are interested in the well known cross covariance function, \citet{Papoulis}, defined as

\begin{equation}
\Xi\left(d_{1}\left(t_{\rm f_{1}}\right),d_{2}\left(t_{\rm f_{2}}\right)\right)= \left.
\frac{\partial^{2}G_{\rm S}}{\partial s_{1}\partial s_{2}} \right\vert_{s_{1},s_{2}=1} -
\left. \frac{\partial G_{\rm S}}{\partial s_{1}} \right\vert_{s_{1},s_{2}=1} \times
\left. \frac{\partial G_{\rm S}}{\partial s_{2}} \right\vert_{s_{1},s_{2}=1},
\label{Cov1}
\end{equation} where $G_{\rm S}\equiv G_{\rm S}\left(s_{1},s_{2}|d_{1}\left(t_{\rm f_{1}}\right),d_{2}\left(t_{\rm
f_{2}}\right)\right)$ is given by (\ref{pgfxi}) with $\xi=0$, i.e., it is the difference
between the cross second factorial moment of the number of detector counts gathered by
both detectors and the product of their first factorial moments. It can be recast as

\begin{equation}
\Xi\left(d_{1}\left(t_{\rm f_{1}}\right),d_{2}\left(t_{\rm f_{2}}\right)\right)= \langle
S_{\rm D_{1}D_{2}}^{+}|\phi\rangle \left(d_{1}\left(t_{\rm
f_{1}}\right),d_{2}\left(t_{\rm f_{2}}\right)\right) + \Delta \Xi \left(d_{1}\left(t_{\rm
f_{1}}\right),d_{2}\left(t_{\rm f_{2}}\right)\right), \label{Cov2}
\end{equation} where the first term takes into account the contribution coming from
multiplicative processes within the system and the detector volumes due to fission events
and detections. The latter can be {\it de facto} neglected if we admit that the volume
occupied by detectors is small in comparison with the system volume. In addition, the
second term in (\ref{Cov2}) stems from the non-Poissonian behaviour of both neutron
sources.

Next we can divide equation (\ref{Cov2}) by $\tau_{\rm c_{1}}\tau_{\rm c_{2}}$ and then apply
the limits $\lim_{\tau_{\rm c_{1}},\tau_{\rm c_{2}}\downarrow 0}$ in order to derive the
expression corresponding to the second order instantaneous rate of the cross covariance function:

\begin{equation}
\ddot{\Xi}\left(t-t_{\rm f_{1}},t-t_{\rm f_{2}}\right)= \langle \ddot{S}_{\rm
D_{1}D_{2}}^{+}|\phi\rangle \left(t-t_{\rm f_{1}},t-t_{\rm f_{2}}\right) + \Delta
\ddot{\Xi} \left(t-t_{\rm f_{1}},t-t_{\rm f_{2}}\right), \label{Cov3}
\end{equation} where, by definition,

\begin{equation}
\ddot{\Xi}\left(t-t_{\rm f_{1}},t-t_{\rm f_{2}}\right)= \lim_{\tau_{\rm c_{1}},\tau_{\rm
c_{2}}\downarrow 0}\frac{\Xi\left(d_{1}\left(t_{\rm f_{1}}\right),d_{2}\left(t_{\rm
f_{2}}\right)\right)}{\tau_{\rm c_{1}}\tau_{\rm c_{2}}}, \label{Xilim}
\end{equation}

\begin{equation}
\Delta\ddot{\Xi}\left(t-t_{\rm f_{1}},t-t_{\rm f_{2}}\right)= \lim_{\tau_{\rm
c_{1}},\tau_{\rm c_{2}}\downarrow 0}\frac{\Delta\Xi\left(d_{1}\left(t_{\rm
f_{1}}\right),d_{2}\left(t_{\rm f_{2}}\right)\right)}{\tau_{\rm c_{1}}\tau_{\rm c_{2}}},
\label{DeltaXilim}
\end{equation} with

\begin{eqnarray}
\langle \ddot{S}_{\rm D_{1}D_{2}}^{+}|\phi\rangle &&  \left(t-t_{\rm f_{1}},t-t_{\rm
f_{2}}\right) \nonumber \\ & =& \langle \ddot{S}_{\rm D_{1}D_{2}}^{+}|\phi^{\rm sf}\rangle
\left(t-t_{\rm f_{1}},t-t_{\rm f_{2}}\right) + \langle \ddot{S}_{\rm
D_{1}D_{2}}^{+}|\phi^{\rm sp}\rangle \left(t-t_{\rm f_{1}},t-t_{\rm f_{2}}\right)
\nonumber
\\& =& \sum_{j,k,l} \bar{\nu}^{2} D \left(\Sigma_{\rm f}\varphi_{j}, \varphi_{
k}^{+},\varphi_{l}^{+}\right)  \int_{-\infty}^{t_{\rm f}} {\rm d}t
\zeta_{j}\left(t\right) \zeta_{{\rm D}_{1} k}^{+}\left(t-t_{\rm f_{1}}\right) \zeta_{{\rm
D}_{2} l}^{+}\left(t-t_{\rm f_{2}}\right), \label{Covsys1}
\end{eqnarray} $D=\overline{\nu\left(\nu-1\right)}/\bar{\nu}^{2}$ being system Diven's
factor, $\bar{\nu}=\sum_{1}^{I}j \varepsilon _{j}$,
$\overline{\nu\left(\nu-1\right)}=\sum_{2}^{I}j (j-1) \varepsilon _{j}$, and where we
have defined the phase-space inner product

\begin{eqnarray}
\left( \Sigma _{\rm f}\varphi _{j}, \varphi _{k}^{+},\varphi _{l}^{+}\right)
&=&\int_{V_{\rm SYS} +V_{\rm D}} \mathrm{d}\mathbf{r}\int \mathrm{d}v\int
\mathrm{d}\mathbf{\Omega }\Sigma_{\rm f}\left( \mathbf{r},v\right) \varphi_{j} \left(
\mathbf{r},v,\mathbf{\Omega }\right) \nonumber \\ &\times& \int \mathrm{d}v^{\prime }\int
\mathrm{d}\mathbf{\Omega }^{\prime } \frac{\chi \left(\mathbf{r}, v^{\prime}
\right)}{4\pi} \varphi_{k}^{+} \left(\mathbf{r}, v^{\prime},\mathbf{\Omega }^{\prime
}\right) \nonumber \\ &\times& \int \mathrm{d}v^{\prime\prime }\int
\mathrm{d}\mathbf{\Omega }^{\prime\prime } \frac{\chi \left(\mathbf{r},
v^{\prime\prime}\right)}{4\pi} \varphi_{l}^{+}\left(\mathbf{r},v^{\prime
\prime},\mathbf{\Omega }^{\prime \prime}\right), \label{inprodfSYS}
\end{eqnarray} and

\begin{eqnarray}
\Delta \ddot{\Xi} \left(t-t_{\rm f_{1}},t-t_{\rm f_{2}}\right) =\Delta \ddot{\Xi}^{\rm
sf} \left(t-t_{\rm f_{1}},t-t_{\rm f_{2}}\right) +\Delta \ddot{\Xi}^{\rm sp}
\left(t-t_{\rm f_{1}},t-t_{\rm f_{2}}\right)  , \label{Covdelta}
\end{eqnarray} which arises from the non-Poissonian nature of the intrinsic and the
external spallation sources, respectively,

\begin{eqnarray}
\Delta \ddot{\Xi}^{\rm sf} && \left(t-t_{\rm f_{1}},t-t_{\rm f_{2}}\right) \nonumber \\ && = \sum_{j,k}
\bar{\nu}_{\rm sf}^{2} D_{\rm sf} \left(S_{1}^{\rm sf},\varphi_{j}^{+},
\varphi_{k}^{+}\right) \int_{-\infty}^{t_{\rm f}} {\rm d}t \zeta_{{\rm D}_{1}
j}^{+}\left(t-t_{\rm f_{1}}\right) \zeta_{{\rm D}_{2} k}^{+}\left(t-t_{\rm f_{2}}\right),
\label{Covdeltasp1}
\end{eqnarray}

\begin{eqnarray}
\Delta \ddot{\Xi}^{\rm sp}&& \left(t-t_{\rm f_{1}},t-t_{\rm f_{2}}\right) \nonumber \\ && = \sum_{j,k}
\left[ \bar{\nu}_{\rm sp} D_{\rm sp} \left(S_{1}^{\rm sp},\varphi_{j}^{+},\varphi_{k}
^{+}\right) + \left(D_{\rm pp} -1\right)\left(S_{1}^{\rm sp},\varphi_{j} ^{+}\right)
\left(S_{1}^{\rm sp},\varphi_{k}^{+}\right) \right] \nonumber
\\ && \times \int_{-\infty}^{t_{\rm f}} {\rm d}t \sum_{m=-\infty}^{\infty}
\delta\left(t-mT\right) \zeta_{{\rm D}_{1} j}^{+}\left(t-t_{\rm f_{1}}\right) \zeta_{{\rm
D}_{2} k}^{+}\left(t-t_{\rm f_{2}}\right), \label{Covdeltasf1}
\end{eqnarray} where $D_{\rm w}=\overline{\nu_{\rm w}\left(\nu_{\rm w}-1\right)}/\bar{\nu} _{\rm
w}^{2}$, ${\rm w}={\rm sf,sp,pp}$, is Diven's factor for the spontaneous fission
(intrinsic) source, the spallation neutron production source, and the pulsed proton
source, respectively, with $\overline{\nu_{\rm w}\left(\nu_{\rm w}-1\right)}=\sum_{2}^{I}j (j-1)
\varepsilon _{j}^{\rm w}$. Furthermore, in the last two expressions we have introduced the
following inner products:

\begin{eqnarray}
\left( S_{1}^{\rm sf},\varphi _{j}^{+},\varphi _{k}^{+}\right) &=& \bar{\nu}_{\rm sf}
N_{0} \int \mathrm{d}\mathbf{r} \rho_{\rm sf}\left(\mathbf{r}\right) \nonumber \\
&\times& \int \mathrm{d}v^{\prime}\int \mathrm{d}\mathbf{\Omega }^{\prime}
\frac{\chi_{\rm S}\left(v^{\prime}\right)}{4\pi} \varphi_{j}^{+} \left(
\mathbf{r},v^{\prime},\mathbf{\Omega }^{\prime}\right) \nonumber \\ &\times& \int
\mathrm{d}v^{\prime\prime}\int \mathrm{d}\mathbf{\Omega }^{\prime\prime} \frac{\chi_{\rm
S}\left(v^{\prime\prime}\right)}{4\pi} \varphi_{k}^{+} \left(
\mathbf{r},v^{\prime\prime},\mathbf{\Omega }^{\prime\prime}\right), \label{inprodsf}
\end{eqnarray}

\begin{eqnarray}
\left( S_{1}^{\rm sp},\varphi _{j}^{+},\varphi _{k}^{+}\right) &=& \bar{\nu}_{\rm pp}
\bar{\nu}_{\rm sp} \int \mathrm{d}\mathbf{r} \rho_{\rm sp}\left(\mathbf{r}\right)
\nonumber \\ &\times& \int \mathrm{d}v^{\prime}\int \mathrm{d}\mathbf{\Omega }^{\prime}
f_{\rm sp}\left(v^{\prime},\mathbf{\Omega }^{\prime}\right) \varphi_{j}^{+} \left(
\mathbf{r},v^{\prime},\mathbf{\Omega }^{\prime}\right) \nonumber \\ &\times& \int
\mathrm{d}v^{\prime\prime}\int \mathrm{d}\mathbf{\Omega }^{\prime\prime} f_{\rm
sp}\left(v^{\prime\prime},\mathbf{\Omega }^{\prime\prime}\right) \varphi_{k}^{+} \left(
\mathbf{r},v^{\prime\prime},\mathbf{\Omega }^{\prime\prime}\right). \label{inprodsp}
\end{eqnarray}

Without loss of generality, we can assume that the upper integral limit in (\ref{Cov1})
can be selected in such a way that $t_{\rm f}>\max\left\{t_{\rm f_{1}},t_{\rm f_{2}}\right\}$, and
then admit that $t_{\rm f}\rightarrow \infty$. Next, we can define the time delay between
the final instant of both detector intervals as $\tau= t_{\rm f_{2}}-t_{\rm f_{1}}$, and
then apply the operator $\int {\rm d}\tau \exp\left(-{\rm i}\omega\tau \right)$ to equation
(\ref{Cov3}) to derive the Fourier transform of $\ddot{\Xi}\left(t-t_{\rm f_{1}},t-t_{\rm
f_{2}}\right)$, i.e., the cross power spectral density:

\begin{equation}
{\rm CPSD}=\mathcal{F}\left[\left\langle \ddot{S}_{\rm D_{1}D_{2}}^{+}|\phi\right\rangle
\right] + \mathcal{F}\left[\Delta\ddot{\Xi} \right], \label{cpsddpm}
\end{equation} where, on account of equations (\ref{zetj}) and (\ref{zeti+}) for $\xi=0$,

\begin{equation}
\mathcal{F}\left[\left\langle \ddot{S}_{\rm D_{1}D_{2}}^{+}|\phi\right\rangle \right] =
\mathcal{F}\left[\left\langle \ddot{S}_{\rm D_{1}D_{2}}^{+}|\phi^{\rm sf}\right\rangle
\right] + \mathcal{F}\left[\left\langle \ddot{S}_{\rm D_{1}D_{2}}^{+}|\phi^{\rm
sp}\right\rangle \right], \label{cpsddpmsys1}
\end{equation}

\begin{eqnarray}
\mathcal{F}\left[\left\langle \ddot{S}_{\rm D_{1}D_{2}}^{+}|\phi^{\rm sf}\right\rangle
\right] &=& \sum_{j,k,l} \bar{\nu}^{2} D \left(\Sigma_{\rm f}\varphi_{j},
\varphi_{k}^{+},\varphi_{l}^{+}\right) \frac{\left(S_{1}^{\rm
sf},\varphi_{j}^{+}\right)}{\left(\frac{1}{v}\varphi_{j},\varphi_{j}^{+}\right)}
\nonumber
\\ &\times& \frac{\left(S_{\rm
D_{1}}^{+},\varphi_{k}\right)}{\left(\frac{1}{v}\varphi_{k}^{+},\varphi_{k}\right)}
\frac{\left(S_{\rm
D_{2}}^{+},\varphi_{l}\right)}{\left(\frac{1}{v}\varphi_{l}^{+},\varphi_{l}\right)}
\frac{\lambda_{\rm sf}}{\left(-\alpha_{j}\right)\left(\omega-{\rm i}\alpha_{k}\right)
\left(\omega +{\rm i} \alpha_{l}\right)},\label{cpsddpmsyssf1}
\end{eqnarray}

\begin{eqnarray}
\mathcal{F}\left[\left\langle \ddot{S}_{\rm D_{1}D_{2}}^{+}|\phi^{\rm sp}\right\rangle
\right] &=& \sum_{j,k,l} \bar{\nu}^{2} D \left(\Sigma_{\rm f}\varphi_{j},
\varphi_{k}^{+},\varphi_{l}^{+}\right) \frac{\left(S_{1}^{\rm
sp},\varphi_{j}^{+}\right)}{\left(\frac{1}{v}\varphi_{j},\varphi_{j}^{+}\right)}
\frac{\left(S_{\rm
D_{1}}^{+},\varphi_{k}\right)}{\left(\frac{1}{v}\varphi_{k}^{+},\varphi_{k}\right)}
\frac{\left(S_{\rm
D_{2}}^{+},\varphi_{l}\right)}{\left(\frac{1}{v}\varphi_{l}^{+},\varphi_{l}\right)}
\nonumber
\\ &\times& \frac{\left[\left(\exp\left(-\alpha_{j}T\right)-1\right)^{-1}-
\left(\exp\left(-\left(\alpha_{k}+{\rm i}\omega\right)T\right)-1\right)^{-1}\right]}
{\left(\omega+{\rm i}\left(\alpha_{j}-\alpha_{k}\right)\right)\left(\omega+{\rm
i}\alpha_{l}\right)}, \label{cpsddpmsyssp1}
\end{eqnarray} where, owing to the application of the deterministic pulsing method, we just
need to assume that $t_{\rm f_{1}}=I_{\rm P}T$, with $I_{\rm P}\rightarrow \infty$,
whereas

\begin{equation}
\mathcal{F}\left[\Delta\ddot{\Xi}\right]= \mathcal{F}\left[\Delta \ddot{\Xi}^{\rm sf}
\right]+\mathcal{F}\left[\Delta \ddot{\Xi}^{\rm sp}\right], \label{DeltaXidpm}
\end{equation}

\begin{eqnarray}
\mathcal{F}\left[\Delta \ddot{\Xi}^{\rm sf}\right] &=& \sum_{j,k} \bar{\nu}_{\rm sf}^{2}
D_{\rm sf} \left(S_{1}^{\rm sf}, \varphi_{j}^{+},\varphi_{k}^{+}\right)
\frac{\left(S_{\rm
D_{1}}^{+},\varphi_{j}\right)}{\left(\frac{1}{v}\varphi_{j}^{+},\varphi_{j}\right)}
\frac{\left(S_{\rm
D_{2}}^{+},\varphi_{k}\right)}{\left(\frac{1}{v}\varphi_{k}^{+},\varphi_{k}\right)}
\nonumber
\\ &\times& \frac{\lambda_{\rm sf}}{\left(\omega-{\rm
i}\alpha_{j}\right)\left(\omega+{\rm i}\alpha_{k}\right)}, \label{DeltaXidpmsf}
\end{eqnarray}

\begin{eqnarray}
\mathcal{F}\left[\Delta \ddot{\Xi}^{\rm sp}\right] &=& \sum_{j,k} \left[ \bar{\nu}_{\rm
sp} D_{\rm sp} \left(S_{1}^{\rm sp},\varphi_{j}^{+},\varphi_{k}^{+}\right) + \left(D_{\rm
pp} -1\right)\left(S_{1}^{\rm sp},\varphi_{j}^{+}\right) \left(S_{1}^{\rm
sp},\varphi_{k}^{+}\right) \right] \nonumber
\\ &\times& \frac{\left(S_{\rm
D_{1}}^{+},\varphi_{j}\right)}{\left(\frac{1}{v}\varphi_{j}^{+},\varphi_{j}\right)}
\frac{\left(S_{\rm
D_{2}}^{+},\varphi_{k}\right)}{\left(\frac{1}{v}\varphi_{k}^{+},\varphi_{k}\right)}%
\frac{\left(\exp\left(-\left(\alpha_{j} +{\rm i} \omega
\right)T\right)-1\right)^{-1}}{\left({\rm i}\omega-\alpha_{k}\right)}.
\label{DeltaXidpmsp}
\end{eqnarray}


\subsection{The stochastic pulsing method}

Similarly, expressions for the factorial moments of the detector number of counts can be
derived for the stochastic pulsing method, just applying Bartlett's procedure to
expression (\ref{pgf}). Now the cross covariance function will be defined as

\begin{eqnarray}
\left\langle\Xi\left(d_{1}\left(t_{\rm f_{1}}\right),d_{2}\left(t_{\rm
f_{2}}\right)\right)\right\rangle_{\xi}&=& \left\langle\left. \frac{\partial^{2}G_{\rm
S}}{\partial s_{1}\partial s_{2}} \right\vert_{s_{1},s_{2}=1}\right\rangle_{\xi} \nonumber \\ & & -
\left\langle \left. \frac{\partial G_{\rm S}}{\partial s_{1}}
\right\vert_{s_{1},s_{2}=1}\right\rangle_{\xi} \times \left\langle\left. \frac{\partial
G_{\rm S}}{\partial s_{2}} \right\vert_{s_{1},s_{2}=1}\right\rangle_{\xi},
\label{Covspm1}
\end{eqnarray} where $G_{\rm S}\equiv G_{\rm S}\left(s_{1},s_{2},\xi|d_{1}\left(t_{\rm f_{1}}
\right),d_{2}\left(t_{\rm f_{2}} \right)\right)$ is given by (\ref{pgfxi}), which can be
recast as

\begin{eqnarray}
\left\langle\Xi \left(d_{1}\left(t_{\rm f_{1}}\right),d_{2}\left(t_{\rm
f_{2}}\right)\right)\right\rangle_{\xi} &=& \left\langle\left\langle
\bar{z}_{1}|S_{1}^{\rm sp}\right\rangle \left\langle \bar{z}_{2}|S_{1}^{\rm sp}
\right\rangle\right\rangle_{\xi} - \left\langle \left\langle \bar{z}_{1}|S_{1}^{\rm
sp}\right\rangle\right\rangle_{\xi} \left\langle\left\langle \bar{z}_{2}|S_{1}^{\rm
sp}\right\rangle\right\rangle_{\xi} \nonumber \\ && +\left\langle
\overline{z_{1}z_{2}}|S_{1}\right\rangle_{\xi}+\left\langle\Delta\Xi\right\rangle_{\xi},
\label{Covspm2}
\end{eqnarray} where $S_{1}^{\rm sp}\equiv S_{1}^{\rm sp}\left(\vartheta,t\right)$ is given by the
second term of expression (\ref{S1}). The first and second term of (\ref{Covspm2}) stem
from the time correlation introduced by the stochastic pulsing
method (see \citet{Ceder2003,BM2005}) and, obviously it only involves the pulsed neutron source.

The cross power spectral density shall be derived following the same steps as before,
i.e., dividing equation (\ref{Covspm2}) by $\tau_{\rm c_{1}}\tau_{\rm c_{2}}$, then applying
the limits $\lim_{\tau_{\rm c_{1}},\tau_{\rm c_{2}}\downarrow 0}$ to the expression
obtained, and, finally, calculating its Fourier transform. Thus, we shall write, on
account of the commutation relations (\ref{comreln1}), (\ref{comreln2}),

\begin{eqnarray}
\left\langle {\rm CPSD}\right\rangle_{\xi} &=& \mathcal{F}\left[\left\langle\left\langle
\dot{S}_{\rm D_{1}}^{+}|\phi^{\rm sp}\right\rangle \left\langle \dot{S}_{\rm
D_{2}}^{+}|\phi^{\rm sp} \right\rangle\right\rangle_{\xi}\right] -
\mathcal{F}\left[\left\langle \left\langle \dot{S}_{\rm D_{1}}^{+}|\phi^{\rm
sp}\right\rangle\right\rangle_{\xi} \left\langle\left\langle \dot{S}_{\rm
D_{2}}^{+}|\phi^{\rm sp}\right\rangle\right\rangle_{\xi}\right] \nonumber \\ &&
+\mathcal{F}\left[\left\langle \ddot{S}_{\rm D_{1}D_{2}}^{+}|\phi
\right\rangle_{\xi}\right]+\mathcal{F}\left[\left\langle\Delta\ddot{\Xi}\right\rangle_{\xi}
\right] , \label{cpsdspm1}
\end{eqnarray} where, as before,

\begin{equation}
\mathcal{F}\left[\left\langle\langle \ddot{S}_{\rm D_{1}D_{2}}^{+}|\phi\rangle
\right\rangle_{\xi}\right] = \mathcal{F}\left[\left\langle \ddot{S}_{\rm
D_{1}D_{2}}^{+}|\phi^{\rm sf}\right\rangle \right] +
\mathcal{F}\left[\left\langle\left\langle \ddot{S}_{\rm D_{1}D_{2}}^{+}|\phi^{\rm
sp}\right\rangle\right\rangle_{\xi} \right], \label{cpsdspmsys1}
\end{equation}

\begin{equation}
\mathcal{F}\left[\left\langle\Delta\ddot{\Xi}\right\rangle_{\xi}\right]=
\mathcal{F}\left[\Delta \ddot{\Xi}^{\rm sf} \right]+\mathcal{F}\left[\left\langle\Delta
\ddot{\Xi}^{\rm sp}\right\rangle_{\xi}\right]. \label{DeltaXispm}
\end{equation} The external pulsed source contribution to the term arising
from the system cross covariance, equation (\ref{cpsdspmsys1}), under the stochastic pulsing
method becomes

\begin{eqnarray}
\mathcal{F}\left[\left\langle\left\langle \ddot{S}_{\rm D_{1}D_{2}}^{+}|\phi^{\rm
sp}\right\rangle\right\rangle_{\xi} \right] &=& \sum_{j,k,l} \bar{\nu}^{2} D
\left(\Sigma_{\rm f}\varphi_{j}, \varphi_{k}^{+},\varphi_{l}^{+}\right)
\frac{\left(S_{1}^{\rm
sp},\varphi_{j}^{+}\right)}{\left(\frac{1}{v}\varphi_{j},\varphi_{j}^{+}\right)}
\nonumber
\\ &\times& \frac{\left(S_{\rm
D_{1}}^{+},\varphi_{k}\right)}{\left(\frac{1}{v}\varphi_{k}^{+},\varphi_{k}\right)}
\frac{\left(S_{\rm
D_{2}}^{+},\varphi_{l}\right)}{\left(\frac{1}{v}\varphi_{l}^{+},\varphi_{l}\right)}
\frac{T^{-1}}{\left(-\alpha_{j}\right)\left(\omega-{\rm i}\alpha_{k}\right) \left(\omega
+{\rm i} \alpha_{l}\right)}. \label{cpsdspmsyssp1}
\end{eqnarray}

On the other hand, the time correlation term stemming from the stochastic pulsing method
in (\ref{Covspm2}), after dividing by $\tau_{\rm c_{1}}\tau_{\rm c_{2}}$ and calculating
the limits $\lim_{\tau_{\rm c_{1}},\tau_{\rm c_{2}}\downarrow 0}$, is given
by, \citet{BM2005},

\begin{eqnarray}
&&\left\langle\left\langle \dot{S}_{\rm D_{1}}^{+}|\phi^{\rm sp}\right\rangle \left\langle
\dot{S}_{\rm D_{2}}^{+}|\phi^{\rm sp} \right\rangle\right\rangle_{\xi} - \left\langle
\left\langle \dot{S}_{\rm D_{1}}^{+}|\phi^{\rm sp}\right\rangle\right\rangle_{\xi}
\left\langle\left\langle \dot{S}_{\rm D_{2}}^{+}|\phi^{\rm
sp}\right\rangle\right\rangle_{\xi}  \nonumber \\ &&\quad\quad = \sum_{j,k}\left(S_{1}^{\rm
sp},\varphi_{j}^{+}\right)\left(S_{1}^{\rm sp},\varphi_{k}^{+}\right) \frac{1}{T^{2}}
\int_{-\infty}^{t_{\rm f}} {\rm d}t^{\prime}\int_{-\infty}^{t_{\rm f}} {\rm
d}t^{\prime\prime} \zeta_{{\rm D}_{1} j}^{+}\left(t^{\prime}-t_{\rm f_{1}}\right)
\nonumber \\ &&\quad\quad \times \zeta_{{\rm D}_{2} k}^{+} \left(t^{\prime\prime}-t_{\rm
f_{1}}-\tau\right) \sum_{m=-\infty,m\neq 0}^{\infty} \exp\left({\rm
i}\frac{2\pi m}{T}\left(t^{\prime}-t^{\prime\prime}\right)\right), \label{corrspm1}
\end{eqnarray} and applying now the operator $\int{\rm d}\tau\exp\left(-{\rm i}\omega\tau
\right)$, we find that

\begin{eqnarray}
&& \mathcal{F}\left[\left\langle\left\langle \dot{S}_{\rm D_{1}}^{+}|\phi^{\rm
sp}\right\rangle \left\langle \dot{S}_{\rm D_{2}}^{+}|\phi^{\rm sp}
\right\rangle\right\rangle_{\xi}\right] - \mathcal{F} \left[\left\langle \left\langle
\dot{S}_{\rm D_{1}}^{+}|\phi^{\rm sp}\right\rangle\right\rangle_{\xi}
\left\langle\left\langle \dot{S}_{\rm D_{2}}^{+}|\phi^{\rm
sp}\right\rangle\right\rangle_{\xi}\right]  \nonumber \\ &&\quad\quad =
\sum_{j,k}\left(S_{1}^{\rm sp},\varphi_{j}^{+}\right)\left(S_{1}^{\rm
sp},\varphi_{k}^{+}\right) \frac{2\pi}{T^{2}} \frac{\left(S_{\rm
D_{1}}^{+},\varphi_{j}\right)}{\left(\frac{1}{v}\varphi_{j}^{+},\varphi_{j}\right)}
\frac{\left(S_{\rm
D_{2}}^{+},\varphi_{k}\right)}{\left(\frac{1}{v}\varphi_{k}^{+},\varphi_{k}\right)} \nonumber \\ && \quad\quad \times
 \sum_{m=-\infty,m\neq 0}^{\infty}
\frac{\delta\left(\omega-\frac{2\pi m}{T}\right)}{\left(\omega-{\rm
i}\alpha_{j}\right)\left(\omega+{\rm i}\alpha_{k}\right)}. \label{corrspm2}
\end{eqnarray}

We can similarly proceed to calculate the second term appearing in equation
(\ref{DeltaXispm}):

\begin{eqnarray}
\mathcal{F}\left[\left\langle\Delta \ddot{\Xi}^{\rm sp}\right\rangle_{\xi}\right] &=&
\sum_{j,k} \left[ \bar{\nu}_{\rm sp} D_{\rm sp} \left(S_{1}^{\rm
sp},\varphi_{j}^{+},\varphi_{k}^{+}\right) + \left(D_{\rm pp} -1\right)\left(S_{1}^{\rm
sp},\varphi_{j}^{+}\right) \left(S_{1}^{\rm sp},\varphi_{k}^{+}\right) \right] \nonumber
\\ &\times& \frac{\left(S_{\rm
D_{1}}^{+},\varphi_{j}\right)}{\left(\frac{1}{v}\varphi_{j}^{+},\varphi_{j}\right)}
\frac{\left(S_{\rm
D_{2}}^{+},\varphi_{k}\right)}{\left(\frac{1}{v}\varphi_{k}^{+},\varphi_{k}\right)}
\frac{\rm 1}{T} \frac{1}{\left(\omega-{\rm i}\alpha_{j}\right) \left(\omega+{\rm
i}\alpha_{k}\right)}. \label{DeltaXispmsp}
\end{eqnarray} This term can effectively become negative for certain
experimental conditions, \citet{BMK2005}.


\section{Fundamental mode approximation and discussion\label{discussion}}

Following the derivation obtained in the previous Section, we can also apply the
fundamental mode approach to the expression corresponding to the CPSD when the stochastic
pulsing method is used:

\begin{equation}
\left\langle {\rm CPSD}\right\rangle_{\xi} =  \left(\varrho_{1} +
\varrho_{2} \sum_{m=-\infty,m\neq 0}^{\infty}
\delta\left(\omega-\frac{2\pi m}{T}\right) \right)
\frac{1}{\left(\omega^{2}+\alpha_{0}^{2}\right) }, \label{cpsdspmfund1}
\end{equation}

\begin{eqnarray}
\varrho_{1} &=& \left\{ \bar{\nu}^{2} D \left(\Sigma_{\rm f}\varphi_{0},
\varphi_{0}^{+},\varphi_{0}^{+}\right) \frac{\left(S_{1}^{\rm sf+sp}
,\varphi_{0}^{+}\right)}{\left(\frac{1}{v}\varphi_{0},\varphi_{0}^{+}\right)}
\frac{1}{\left(-\alpha_{0}\right)} + \bar{\nu}_{\rm sf}^{2} D_{\rm sf} \left(S_{1}^{\rm
sf}, \varphi_{0}^{+},\varphi_{0}^{+}\right) \right. \nonumber \\ && + \left. \frac{\rm
1}{T} \left[ \bar{\nu}_{\rm sp} D_{\rm sp} \left(S_{1}^{\rm
sp},\varphi_{0}^{+},\varphi_{0}^{+}\right) + \left(D_{\rm pp} -1\right)\left(S_{1}^{\rm
sp},\varphi_{0}^{+}\right) \left(S_{1}^{\rm sp},\varphi_{0}^{+}\right) \right] \right\}
\nonumber
\\ &\times&  \frac{\left(S_{\rm
D_{1}}^{+},\varphi_{0}\right)}{\left(\frac{1}{v}\varphi_{0}^{+},\varphi_{0}\right)}
\frac{\left(S_{\rm
D_{2}}^{+},\varphi_{0}\right)}{\left(\frac{1}{v}\varphi_{0}^{+},\varphi_{0}\right)} ,
\label{rho1} \\%
\varrho_{2}  &=& \left(S_{1}^{\rm sp},\varphi_{0}^{+}\right)\left(S_{1}^{\rm
sp},\varphi_{0}^{+}\right) \frac{2\pi}{T^{2}} \frac{\left(S_{\rm
D_{1}}^{+},\varphi_{0}\right)}{\left(\frac{1}{v}\varphi_{0}^{+},\varphi_{0}\right)}
\frac{\left(S_{\rm
D_{2}}^{+},\varphi_{0}\right)}{\left(\frac{1}{v}\varphi_{0}^{+},\varphi_{0}\right)} .
\label{rho2}
\end{eqnarray} where we have defined the total neutron source strength $S_{1}^{\rm
sf+sp}\equiv S_{1}^{\rm sf+sp} \left(\vartheta,t \right)=\lambda_{\rm sf} S_{1}^{\rm sf}
\left(\vartheta,t \right)+T^{-1}S_{1}^{\rm sp} \left(\vartheta,t \right)$.

Equation (\ref{cpsdspmfund1}) apparently seems to be similar to that obtained by
\citet{RKW} albeit a deeper examination permits to understand an important
difference: in the latter case, the expression for the CPSD for the stochastic pulsing
method was derived assuming a quasi-Poissonian behaviour of the pulsed external source,
i.e., with no delay time averaging. That means that the expression obtained in this reference
should be exactly equal to our expression for the CPSD in Section \ref{cpsdsec} under the
deterministic pulsing method, only without the contribution of (\ref{DeltaXidpmsp}),
which stems from the non-Poissonian behaviour of the periodic pulsed source. But it is
clear that, in any case, equation (\ref{cpsddpm}) is a completely bounded function for any non-zero value of
the frequency, $\omega$, which cannot produce the response obtained at MUSE-4
experiments, also reported by \citet{RKW}, for the stochastic pulsing CPSD
method. Unlikely, we have shown that those spectral lines appearing at frequencies which
are multiples of the accelerator frequency are indeed not produced by the system cross
covariance contribution, but they are a {\it prima facie} of the time self-correlation
introduced by the stochastic pulsing method.

In addition, it has been recently shown that the utilisation of deterministic external
pulsed sources, such as those used in MUSE-4 experiments, can make the nuclear process to
behave as a sub-Poissonian one, \citet{BMK2005}. It can occur when the contribution of
non-Poissonian term of the pulsed source becomes negative. That means that under those
conditions the time-independent term $\varrho_{1}$ can be effectively negative for very
deterministic pulsed sources. But, anyway, it is of common practice to use Bode's diagram
of the magnitude in order to represent graphically a system response function such as
(\ref{cpsdspmfund1}), therefore, it is not necessary to consider the sign of
$\varrho_{1}$ provided that, from a practical viewpoint, we only need to fit its
magnitude for those points of the graph where $\omega$ is not a multiple of the
accelerator frequency, together with the eigenvalue $\alpha_{0}$. Notice that the value
of $\varrho_{2}$ cannot be determined from this technique (we will need a further
integral condition), but in any case it is completely useless for a practical purpose.
That means that when the stochastic pulsing CPSD method is applied to determine the value
of the subcriticality level of a nuclear system, only two fitting parameters must be
considered, in contrast with the stochastic pulsing Feynman-$\alpha$ technique, where three parameters must be fitted, \citet{BMK2005}.

\begin{figure}
\centering{
\includegraphics{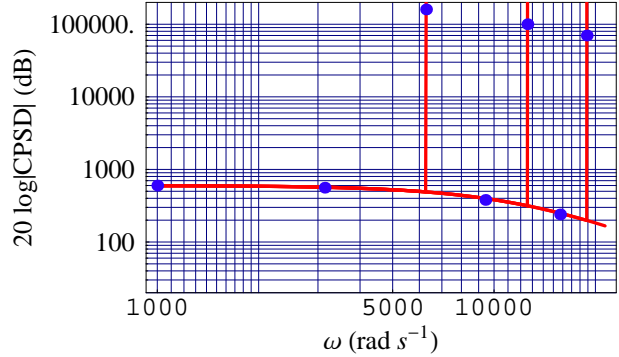}
\caption{\label{fig:cpsd} Stochastic CPSD pulsing method obtained during MUSE-4
experiments for the SC0 configuration of the MASURCA subcritical assembly. Points of the
experimental curve (points) from \citet{RKW} are compared with equation
(\ref{cpsdspmfund1}) fitted with $(-\alpha_{0})=13258\pm273$ $rad\cdot s^{-1}$ (solid).
The accelerator period is equal to $1$ $ms$.}}
\end{figure}

In Figure \ref{fig:cpsd} we show some experimental points reported in that graph
corresponding to the stochastic pulsing CPSD method of \citet{RKW} (Figure 6),
together with equation (\ref{cpsdspmfund1}) conveniently fitted. This particular experiment,
corresponds to the configuration SC0 of the MASURCA subcritical assembly used during
MUSE-4 studies, \citet{Soule}, with a D-D pulsed source. The value of the prompt neutron
time constant obtained by \citet{RKW}, which is shown in Table \ref{table}, is
effectively very similar to that reported using other noise techniques for the same
conditions.

\begin{table}[t]
\caption{\label{table}Values of $(-\alpha_{0})$ obtained for the configuration SC0 (with
pilot rod inserted) of the MASURCA subcritical assembly used during MUSE-4
experiments, \citet{Soule}. The result of the eigenvalue fitting by means of the Feynman-$\alpha$ method is reported by \citet{BMK2005}.}
\begin{tabular}{lc}
Method & $(-\alpha_{0})$ $(rad\cdot s^{-1})$\\ \hline%
Stochastic Pulsing CPSD method & 13258$\pm$273\\ %
Stochastic Pulsing Feynman-$\alpha$ method & 13646$\pm$515\\
\end{tabular}
\end{table}


\section{Conclusions}

In the present work we have dealt with the applicability of
stochastic-neutron-field-based methods for the study of the neutron counting statistics
in a nuclear system. We have derived the generalised two-detectors relationship between
the probability generating functions of the kernel and the source for subcritical
assemblies when pulsed neutron sources are used together with the intrinsic neutron
source coming from spontaneous fission events within the fuel material, \citet{BM2005}. It
has been done within the stochastic neutron transport theory framework, which permits to
understand how the general transport problem is influenced by its spatial, spectral and
angular dependence.

Further, we have followed P\'al-Bell's methodology for the derivation of the
integro-differential Boltzmann transport equation, and applied the formalism described by
\citet{MPV87} in order to calculate the chosen statistical descriptor.

In Section \ref{cpsdsec} an expansion in $\alpha$-eigenvalues for the cross covariance
and the CPSD of two-detectors stochastic counting rates have been obtained. The
contribution of higher harmonics in subcritical monitoring problems shall play an
important role in ADS assemblies. In this case, the excitement of higher modes could be
relevant in situations of normal operation, and it will increase as the reactor departs
from the criticality condition.

In Section \ref{discussion} we have compared the expression obtained for the stochastic
pulsing CPSD method with experimental data obtained during the MUSE-4 European project.
The value of the prompt neutron time constant fitted is comparable with others methods.
In addition, the reduced number of fitting parameters makes this method suitable as a
subcriticality monitoring technique.


\section{Acknowledgements}

The work of D. B. has been supported by grants FPU AP2003-3847 from the Spanish Ministry
of Education and Science.

\appendix





\begin{thebibliography}{}



\bibitem[Ballester and Mu\~noz-Cobo(2005)]{BM2005} Ballester, D., Mu\~noz-Cobo, J. L., 2005. Ann. Nucl. Energy {\bf 32}, 493-519.

\bibitem[Ballester {\it et al.}(2005)]{BMK2005} Ballester, D., Mu\~noz-Cobo, J. L., Kloosterman, J. L., 2005. Ann. Nucl. Energy, in press.

\bibitem[Bartlett(1955)]{Bartlett} Bartlett, M. S., 1955. An Introduction to Stochastic Processes, Cambridge University Press, Cambridge, UK.

\bibitem[Bell(1965)]{Bell65} Bell, G. I., 1965. Nucl. Sci. Eng. {\bf21}, 390-401.

\bibitem[Bell and Glasstone(1979)]{Bell79} Bell, G. I., Glasstone, S., 1970. Nuclear Reactor
Theory, Krieger Publishing Co., Malabar, FL.

\bibitem[Behringer and Wydler(1999)]{Behringer99} Behringer, K., Wydler, P., 1999. Ann. Nucl. Energy {\bf 26}, 1131-1157.

\bibitem[Carta and D'Angelo(1999)]{Carta99} Carta, M., D'Angelo, A., 1999. Subcriticality-level evaluation in
accelerator-driven systems by harmonic modulation of the external source. Nucl. Sci. Eng. {\bf 133}, 282-292.

\bibitem[Ceder and P\'{a}zsit(2003)]{Ceder2003} Ceder, M., P\'{a}zsit, I., 2003. Prog. Nucl. Energy {\bf 43}, 429.

\bibitem[Courant and Wallace(1947)]{Courant} Courant, E. D., Wallace, P. R., 1947. Phys. Rev. {\bf72}, 1038-1048.

\bibitem[Degweker(2003)]{Degweker2003} Degweker, S. B., 2003. Ann. Nucl. Energy {\bf 30}, 223-243.

\bibitem[Lando(2003)]{Lando} Lando, S. K., 2003. Lectures on Generating Functions, Amer. Math. Soc., Providence, RI.

\bibitem[Lewins(1978)]{Lewins78} Lewins, J., 1978. Nuclear Reactor Kinetics and Control. Pergamon Press., Oxford.

\bibitem[Matthes {\it et al.}(1988)]{Matthes88} Matthes, W., 1988. Some applications of stochastic processes in neutron coincidence: measurements used in nuclear safeguards, Proc. NATO ARW on Noise and Nonlinear Phenomena in Physical Systems, Ed. Plenum Serie B, {\bf 192}.

\bibitem[Mu\~noz-Cobo {\it et al.}(1987)]{MPV87} Mu\~{n}oz-Cobo, J. L., Perez, R. B., Verd\'{u}, G., 1987. Nucl. Sci. Eng. {\bf 95}, 83-105.

\bibitem[Mu\~noz-Cobo and Verd\'{u}(1987)]{MV87} Mu\~{n}oz-Cobo, J. L., Verd\'{u}, G., 1987. Ann. Nucl. Energy {\bf 14}, 327-350.

\bibitem[Mu\~noz-Cobo {\it et al.}(2000)]{Munoz2000} Mu\~{n}oz-Cobo, J. L., Perez, R. B., Valentine, T. E., Rugama, Y.,
Mihalczo, J. T., 2000. Ann. Nucl. Energy {\bf 27}, 1087-1114.

\bibitem[P\'{a}l(1958)]{Pal58} P\'{a}l, L., 1958.  Il Nuovo Cimento Suppl. VII, {\bf 25}.

\bibitem[Papoulis(1991)]{Papoulis} Papoulis, A., 1991. Probability, Random Variables, and Stochastic Processes, 3rd ed. McGraw-Hill Book Co., Singapore.

\bibitem[Perez {\it et al.}(1964)]{Perez65} Perez, R. B., Booth, R. S., Denning, R. S., Hartley,
R. H., 1964. Trans. Am. Nucl. Soc.
{\bf 7(1)}, 49-50.

\bibitem[Rugama {\it et al.}(2004)]{RKW} Rugama, Y., Kloosterman, J. L., Winkelman, A., 2004. Prog. Nucl. Energy {\bf 44}, 1-12.

\bibitem[Soule {\it et al.}(2004)]{Soule} Soule, R., {\it et al.}, 2004. Nucl. Sci. Eng. {\bf 148}, 124-152.

\bibitem[Uhrig(1970)]{Uhrig} Uhrig, R., 1970. Random Noise Techniques in Nuclear Reactor Systems.
Ronald Press, New York.

\bibitem[Valentine {\it et al.}(2000)]{Valentine2000} Valentine, T. E., Rugama, Y., Mu\~{n}oz-Cobo, J. L., Perez, R. B.,
2000. Coupling of MCNP-DSP and LAHET
codes for designing subcritical monitors for accelerator driven systems. Proc. Monte-Carlo Conference, Lisbon (Portugal),
edited by Springer Verlag.

\bibitem[Williams(1974)]{Williams74} Williams, M. M. R., 1974. Random Processes in
Nuclear Reactors. Pergamon Press., Oxford.

\end{thebibliography}
\end{document}